\documentclass[prl, twocolumn, showpacs, amsmath, amssymb,superscriptaddress]{revtex4-1}
\usepackage{amssymb,amsfonts,amsmath} 
\usepackage{graphicx}% Include figure files 
\usepackage{color}
\usepackage{hyperref}
\usepackage{longtable}
\usepackage{bbm}
\usepackage{ulem}
\usepackage{enumitem}
\usepackage{bm}
\usepackage{braket}
\usepackage{comment}
\usepackage{nicefrac}

\begin{document} 

%\title{ Anderson Quantum Anamolous and Spin Hall States in Magnetically Doped\\ {(Bi, Sb)$_2$Te$_3$} Films}
\title{Topological Phase Transitions Induced by Disorder in Magnetically Doped \\(Bi, Sb)$_2$Te$_3$ Thin Films}

\author{Takuya Okugawa}
%\email{okugawa@physik.rwth-aachen.de}
\affiliation{Institut f\"ur Theorie der Statistischen Physik, RWTH Aachen, 
52056 Aachen, Germany and JARA - Fundamentals of Future Information Technology.}

\author{Peizhe Tang}
\email{peizhet@buaa.edu.cn}
\affiliation{School of Materials Science and Engineering, Beihang University, Beijing 100191, P. R. China.}
\affiliation{Max Planck Institute for the Structure and Dynamics of Matter, Center for Free Electron Laser Science, 22761 Hamburg, Germany.}

\author{Angel Rubio}
\affiliation{Max Planck Institute for the Structure and Dynamics of Matter, Center for Free Electron Laser Science, 22761 Hamburg, Germany.}
\affiliation{Center for Computational Quantum Physics, Simons Foundation Flatiron Institute, New York, NY 10010 USA.}
\affiliation{Nano-Bio Spectroscopy Group, Departamento de Fisica de Materiales, Universidad del Pa\'is Vasco, UPV/EHU- 20018 San Sebasti\'an, Spain.}

\author{Dante M.\ Kennes}
\email{Dante.Kennes@rwth-aachen.de}
\affiliation{Institut f\"ur Theorie der Statistischen Physik, RWTH Aachen, 
52056 Aachen, Germany and JARA - Fundamentals of Future Information Technology.}
\affiliation{Max Planck Institute for the Structure and Dynamics of Matter, Center for Free Electron Laser Science, 22761 Hamburg, Germany.}

\begin{abstract} 
We study disorder induced topological phase transitions in magnetically doped (Bi, Sb)$_2$Te$_3$ thin films, by using large scale transport simulations of the conductance through a disordered region coupled to reservoirs in the quantum spin Hall regime. Besides the disorder strength, the rich phase diagram also strongly depends on the magnetic exchange field, the Fermi level, and the initial topological state in the undoped and clean limit of the films. In an initially trivial system at non-zero exchange field, varying the disorder strength can induce a sequence of transitions from a normal insulating, to a quantum anomalous Hall, then a spin-Chern insulating, and finally an Anderson insulating state. While for a system with topology initially, a similar sequence, but only starting from the quantum anomalous Hall state, can be induced. Varying the Fermi level we find a similarly rich phase diagram, including transitions from the quantum anomalous Hall to the spin-Chern insulating state via a state that behaves as a mixture of a quantum anomalous Hall and a metallic state, akin to recent experimental reports.  
\end{abstract}

\date{\today} 
\maketitle

%\section{Introduction}

\textit{Introduction.}--- The interplay between magnetism and topological states of matter has attracted tremendous research interests in the past decades for its value in both fundamental science and applications. Many interesting and exotic phenomena  have been predicted and observed in these types of  systems, such as antiferromagnetic (AFM) topological insulators (TIs) \cite{Mong2010,otrokov2019prediction,liJH2018}, AFM Dirac semimetals \cite{tang2016dirac}, and magnetic Weyl semimetals \cite{Armitage2018}. The observation of the quantum anomalous Hall (QAH) effect in two dimensional (2D) topological materials, in which a dissipationless quantized Hall conductance carried by chiral edge states is found in transport measurements \cite{yu2010quantized,chang2013experimental,liu2016quantum,he2018topological}, is one of the most important contributions to this field. In contrast to the quantum Hall effect, the QAH effect does not necessitate any external magnetic field, but instead requires the spontaneous formation of long-range magnetic order inside the 2D materials which breaks the time reversal symmetry \cite{ChangCZ2013,yu2010quantized,zhang2013topology}. Currently, the QAH effect has been observed in twisted bilayer graphene \cite{sharpe2019emergent,serlin2020intrinsic}, MnBi$_2$Te$_4$ thin film with an odd number of septuple layers \cite{deng2020quantum}, and magnetically doped TI thin film \cite{yu2010quantized,chang2013experimental,Checkelsky2014Trajectory,chang2015high,FengYang2015,Kou2014,Bestwick2015,Chang2016Obse,Yasuda2017Quantized,mogi2015magnetic}.  

One of the first QAH material candidates, which have been well-studied in the past few years, are  magnetically doped TI thin films, such as Cr or V-doped (Bi, Sb)$_2$Te$_3$ with a thickness of several quintuple layers (QLs) \cite{chang2013experimental,Checkelsky2014Trajectory,FengYang2015,Kou2014,chang2015high}. Here, long-range ferromagnetic (FM) order is achieved by magnetic doping \cite{yu2010quantized,chang2013experimental,ChangCZ2013,zhang2013topology}. Once FM order is formed in these thin films, the strength of exchange field can be tuned efficiently by changing the concentration of magnetic ions and the chemical potential \cite{LiuQ2009,checkelsky2012dirac,ZhuJJ2011,chang2013thin,ChangCZ2014}. Furthermore, it was shown experimentally that electronic and topological properties of  magnetically doped TI thin films can also be manipulated precisely via changing their thickness \cite{He2010Crossover}, tuning the chemical constituents \cite{zhang2013topology}, or applying a dual-gate technology \cite{wang2015,zhang2017magnetic}. Being alloy compounds, TI thin films exhibit disorder naturally, which can be classified into two kinds. The one kind is magnetic disorder, e.g. induced by the magnetic dopands,  that induces  spin flips by scattering and breaks the time reversal symmetry. The other kind is non-magnetic, does not break time reversal symmetry but does induce spatial inhomogeneity. For the first one, there are many recent studies  \cite{Namura2011,LuHZ2011,QiaoZH2016,ChenCZ2019,Haim2019,xing2018influence,WangJ2014Uni,Keser2019,wang2018direct,lee2015imaging,lachman2015vis,kou2015metal}, which conclude that weak magnetic disorder could stabilize a QAH effect and induce new topological phases, while strong disorder will drive the system towards an Anderson insulator state. The second type of disorder, has received less theoretical attention \cite{zhang2020chiral}, although it is well-known that spatial inhomogeneity of thin films is an important factor in experiments \cite{wray2011topological,wang2018direct,lee2015imaging,YuanYH2020,chen2015magnetism,Chang2016Obse,LiaoJ2015}--- a shortcoming that we will remedy within this letter, by focusing on this second type of disorder. This is particularly pressing as for intrinsic HgTe/CdTe quantum wells, this second type of disorder can induce a topological phase transition (TPT) driving the quantum well from a normal insulator (NI) to a topological Anderson insulator (TAI) with quantized edge conductance \cite{Li2009,Groth2009,Jiang2009,Yamakage2013,SongJ2012}. 
%"Long-range FM" order is a standard expression.  

\begin{figure}[t]
{\includegraphics[width=0.8\columnwidth,clip]{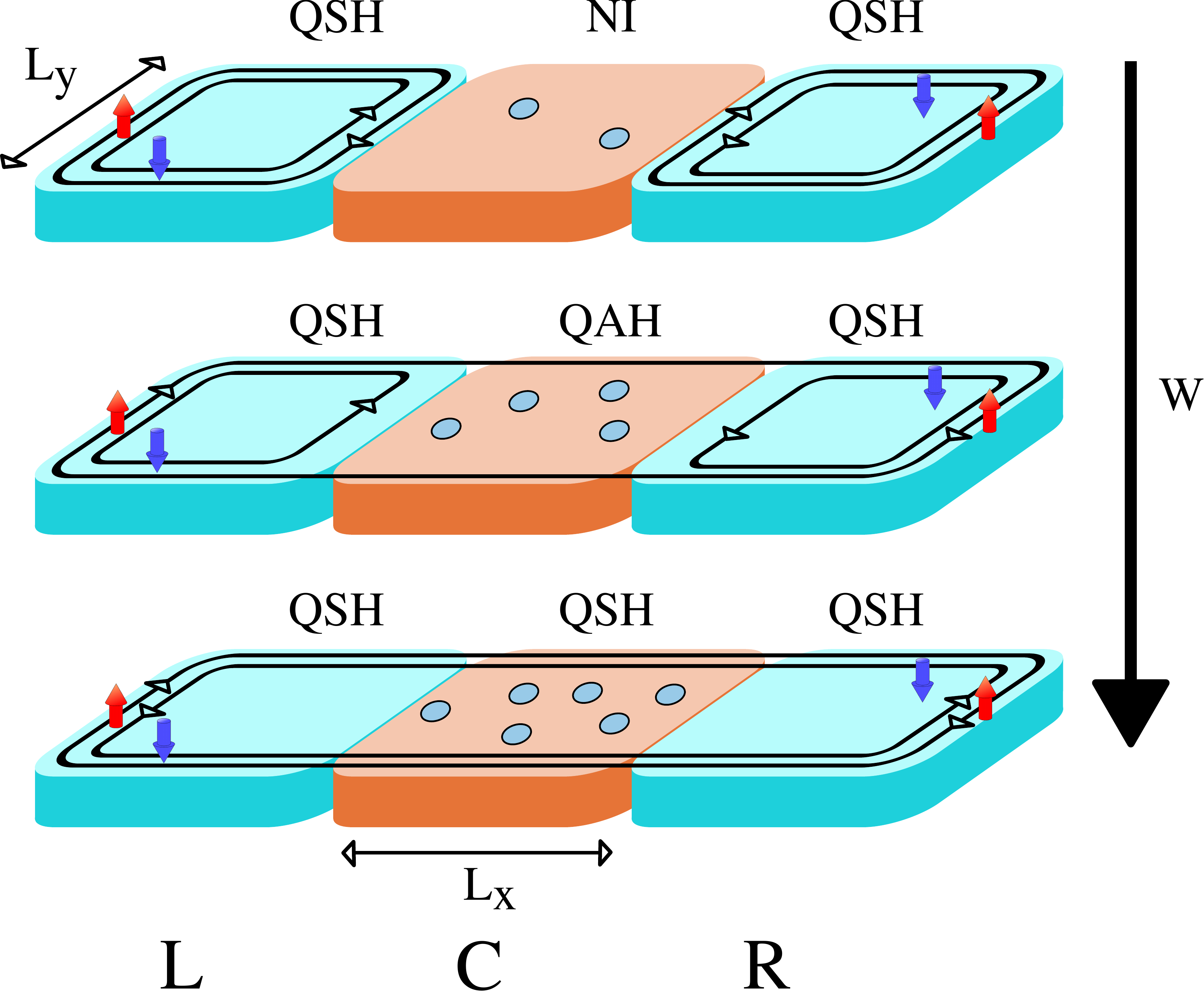}}
\caption{Schematic of the system. The QSH leads are given in light blue, while the central region is colored in orange. The disorder strength can tune a NI to a QAH and even to a QSH state as the disorder is varied. Depending on the phase the conductance is quantized to $0$, $1$ or $2$ times the conductance quantum carried by edge states with locked pseudo-spin.}
\label{system}
\end{figure}

\begin{figure}[t]
{\includegraphics[width=0.9\columnwidth,clip]{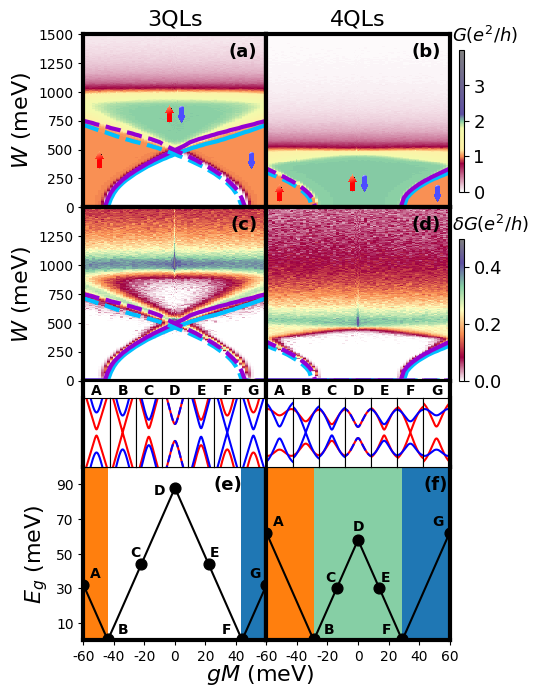}}
%{\includegraphics[width=0.9\columnwidth,clip]{{temp_figure/4QL_metalead_qshi_G_dG_Mex=[0.0, 10.0, 20.0]_EF=-80-119_100_imp_0-1500_ave=500_eta_-3_m0_-29_2}.png}}
\caption{Phase diagram of magnetically doped (Bi, Sb)$_2$Te$_3$ thin film with the thickness of 3QLs and 4QLs, corresponding to left and right column, respectively. (a) and (b) show the average conductance $G$ and (c) and (d) the standard deviation $\delta G$ as a function of the exchange field $gM$ and disorder strength $W$. The system sizes are $L_x=400a$ and $L_y=100a$, $a$ is the lattice constant.
%The other parameters are $E_F=1 $ meV, $v_F=3.07/2.36$ eV \AA, $m_0=44/-29$ meV, $B=37.3/12.9$ eV \AA$^2$ for (a)(c)/(b)(d)~\cite{wang2015}. 
The Fermi level is $1$ meV. The disorder average is taken over 500 random configurations. The red and blue spins indicate the pseudo-spins that distinguish the upper and lower block in Eq.~\eqref{hamiltonian}. The colored lines represent the phase boundaries from self-consistent Born approximation. %~\cite{Note_parameter}. %The phase boundary lines are obtained from the . The blue and purple lines show $|\overline{\mu}^{u/l}|=\overline{m}_{0}^{u/l}$ and $|\overline{\mu}^{u/l}|=-\overline{m}_{0}^{u/l}$, respectively. %which carry corresponding conductance in each phase domain.
Details about the colored lines and the other parameters are given in \cite{Note_parameter}. (e) and (f) show the bulk band-gap as a function of exchange field $gM$. The upper subpanels show bulk band sketches that correspond to the labeled points, the red (blue) lines represent the energy bands from the upper (lower) block in Eq.~\eqref{hamiltonian}. The background color indicates different topological phases. Orange denotes a QAH insulator with Chern number $C=-1$, blue a QAH insulator with $C=+1$, green a spin-Chern insulator, and white a NI.
%, where $C_{+}(C_{-})$ denotes Chern number calculated from the upper(lower) block of the Hamiltonian in eq. (\ref{hamiltonian}).
}
\label{Mex_disorder}
\end{figure}

In this work, by using large scale tight binding transport simulations of disordered 2D systems, we reveal the surprisingly rich phase diagram of magnetically doped TI thin films, which depends on the non-disordered (undoped) parent state in an intricate fashion. A schematic of our simulations is summarized in Fig.~\ref{system}. We use a large central region exhibiting disorder of varying strength coupled to two semi-infinite leads. The transport via edge states through the central region depends on its topological phase and can be flexibly tuned by disorder strength. This finding is illustrated in detail in Fig.~\ref{Mex_disorder} (a) and (b) for (Bi, Sb)$_2$Te$_3$ thin films with the thickness of 3QLs and 4QLs, hosting a NI and quantum spin Hall (QSH) state without disorder and exchange field, respectively. In dependence of the number of layers, disorder strength $W$ and exchange field $gM$, a rich behavior of the topological phases is found by considering the conductance. 
Disorder can induce a series of transitions from NI$\to$ QAH$\to$ spin-Chern insulator $\to$ Anderson insulator as its strength is increased to larger values.
%removed TR but not TPT it is used more often later on.
%{\color{blue}TO: we can remove the abbreviation of TR and TPT because these abbreviations are not used in anywhere except here. Or if we use these, we should use these abbreviations in the other sections of this paper as well}

%\section{Model and method}
\textit{Models and Methods}.--- We aim at modelling the electronic structures of magnetically doped (Bi, Sb)$_2$Te$_3$ thin films around the Fermi level at the $\Gamma$-point by starting from the low-energy $\bm{k}\cdot \bm{p}$ effective Hamiltonian \cite{yu2010quantized}:    
\begin{equation}
H_0=
\begin{pmatrix}
h(\bm{k}) +gM\sigma_z & 0 \\
0 & h^*(\bm{k})-gM\sigma_z \label{hamiltonian} \\
\end{pmatrix},
\end{equation}
%\begin{align}
with $h(\bm{k})=\bm{d(\bm{k})} \cdot \bm{\sigma}$, % \\
$\bm{d(\bm{k})}=(v_F k_y, -v_F k_x, m(\bm{k}))$ and % \\
$m(\bm{k})=m_0+B(k_x^2+k_y^2)$.
%\end{align}
Here, the basis is chosen as $\ket{+ \uparrow}$, $\ket{- \downarrow}$, $\ket{+ \downarrow}$, $\ket{- \uparrow}$, where $\ket{\pm \uparrow (\downarrow)}=(\ket{t \uparrow (\downarrow)} \pm \ket{b \uparrow (\downarrow)})$, and $t (b)$ represents the top (bottom) surface states. $\bm{\sigma}=(\sigma_x,\sigma_y,\sigma_z)^T$ is a vector with the entries being the Pauli matrices for real spin ($\uparrow$ and $\downarrow)$. $v_F$ is the Fermi velocity, $M$ is the exchange field in the z-direction, and $g$ is the effective $g$-factor. $m(\bm{k})$ describes the tunneling between the top and the bottom surface, which goes to zero as the thickness of the film is increased. Since the Hamiltonian shown in Eq.~\eqref{hamiltonian} is block diagonal, topological properties and the energy dispersion can be defined for each block of the Hamiltonian separately~\cite{ShengDN2006}. 
%We define the respective dispersions $E^{u/l}=\pm \sqrt{(v_F k)^2 + m^{u/l} (\bm{k})^2}$, where $m^{u/l} (\bm{k})=m(\bm{k}) \pm gM$. 

In order to quantify the effect of disorder on the magnetic TI, we perform transport calculation as a function of both disorder strength $W$ and the effective exchange field $gM$. We use a lattice version of the Hamiltonian of Eq.~\eqref{hamiltonian} with quenched random on-site disorder uniformly distributed within $[-W/2, W/2]$ (see SM for details~\cite{SuppMater}). We then calculate the disorder averaged conductance $G$ and the corresponding standard deviation $\delta G$ of a stripe geometry using the Landauer-B\"uttiker formula~\cite{Landauer1970, Buttiker1988}. A similar approach was applied to other topological materials showing a TAI phase~\cite{Li2009, Groth2009, Jiang2009}. Our stripe geometry consists of a disordered central region with the length $L_x$ and width $L_y$ connected to a left and right semi-infinite clean lead (see Fig.~\ref{system}). In contrast to previous works we here suggest leads in the QSH regime to probe the conductance. This choice of leads, though not affecting the central region's physics (see SM~\cite{SuppMater}), allows us to probe the conductance much more clearly and we suggest to use a similar setup in future experiments.     
%$M=gM$??? ({\color{red} PZ: we should make the expression in Figures and Euq1 consistent.}). 

We compare our exact simulations on disorder induced TPT with a self-consistent Born approximation~\cite{Groth2009}, in which the effect of disorder can be subsumed in a change of the topological mass term and the chemical potential~(see SM for details~\cite{SuppMater}). Following the two scenarios presented in the original proposal to achieve the QAH effect via magnetic doping of TI thin films \cite{yu2010quantized}, we compare two kinds of (Bi, Sb)$_2$Te$_3$ thin films in this work whose topological properties can be tuned by the quantum confinement \cite{LiuCX2010-Osci,He2010Crossover}: One kind is a trivial thin film with the thickness of 3QLs in the clean and undoped limit, the other is a 4QLs thin film being in the QSH state under the same condition. The doping of magnetic ions, such as Cr and V, will stabilize the long-range FM order in the 2D bulk states \cite{chang2013experimental,chang2015high} and the controllable exchange field $gM$ can drive both of the two kinds of thin films to the QAH phase \cite{yu2010quantized}. In our simulations, one QL is about 1nm thick and we use the effective parameters specific to (Bi, Sb)$_2$Te$_3$ films from Ref.~\cite{wang2015} for the Hamiltonian shown in Eq.~\eqref{hamiltonian}. %We employ these parameters and sweep through exchange field and disorder strength.

\begin{figure}[t]
{\includegraphics[width=0.9\columnwidth,clip]{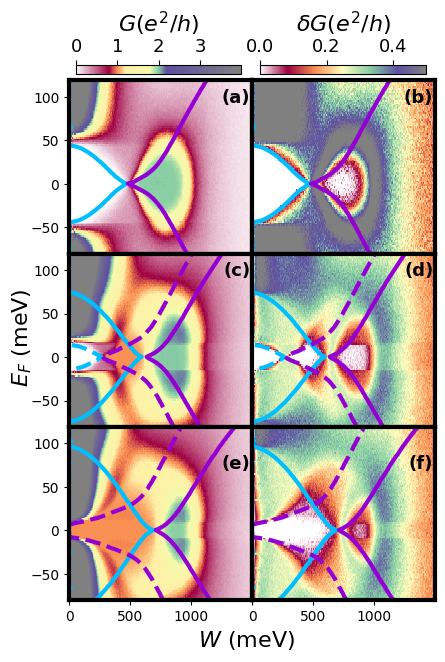}}
\caption{Left (a), (c), (e) and right (b), (d), (f) panels show the average conductance $G$ and the corresponding standard deviation $\delta G$, respectively, as a function of disorder strength $W$ and Fermi energy $E_F$ for 3QLs (Bi, Sb)$_2$Te$_3$ thin film with and without magnetic doping. The exchange fields $gM$ are taken to be (a), (b): 0 meV, (c), (d): 30 meV, and (e), (f): 52 meV. The color lines stand for phase boundaries from self-consistent Born approximation. %~\cite{Note_parameter}. %$v_F=3070$ meV \AA, $m_0=44$ meV, $B=37300$ meV \AA$^2$ for all, which are corresponding to 3QLs~\cite{wang2015}.
%Positive $m_0$ indicates trivial insulator regime. 
Details of the colored lines and calculation parameters are shown in Ref.~\cite{Note_parameter}. The system sizes are $L_x=400a$ for all panels and $L_y=100a$ for (a),(b) and $L_y=200a$ for (c)-(f). The disorder average is taken over 500 random configurations.}
\label{3QL}
\end{figure}

\textit{Disorder Induced TPT in Magnetic Thin Films}.--- Figure~\ref{Mex_disorder} summarizes the phase diagram obtained from the conductance $G$ and the corresponding standard deviation $\delta G$ with respect to exchange field $gM$ and disorder strength $W$ both for magnetically doped TI thin films with thickness of 3QLs and 4QLs when the Fermi energy is located inside the intrinsic band gap. Our work focuses on the effects of disorder, but to paint the full picture we first discuss the $W=0$ limit and then systematically include the disorder. At zero exchange field 3QLs and 4QLs can be viewed as tokens of two generic classes, either featuring a trivial NI or QSH state. As the exchange field is increased, time reversal symmetry is broken. The QSH state turn to a spin-Chern insulator~\cite{ShengDN2006,noteSpinChern}, but the NI remains trivial. At larger exchange fields a transition to a QAH phase is found for both 3QLs and 4QLs. 
%either by inverting one of the formerly trivial bands to a topological one (3QLs) or inverting one topological to a trivial one (4QLs). 
The NI, QAH and QSH or spin-Chern insulators are characterized by a quantized conductance of $G=0$ (white), $G=e^2/h$ (orange) and $G=2e^2/h$ (green), respectively in Fig.~\ref{Mex_disorder} (a) and (b). Quantized conductance reflects in vanishing standard deviations in these regions as shown in Fig.~\ref{Mex_disorder} (c) and (d). Still for clean samples, we show the bulk gap in dependence of exchange field in Fig.~\ref{Mex_disorder} (e) and (f). The bulk bands around the $\Gamma$ point are sketched above as sub-panels. The background color indicates different topological phases. By increasing the exchange field in magnitude, the two cases will host a QAH phase with the same Chern number, but the inverted bands belong to different blocks in the Hamiltonian Eq.~\eqref{hamiltonian}, %(see colored bands, with the change of Chern numbers of upper and lower block $C_{+/-}$), 
because the exchange field either inverts one of the formerly trivial bands to a topological one (3QLs) or inverts one topological to a trivial one (4QLs).

We now turn to finite disorder and first concentrate on $gM=0$. For the 3QL case we find a phase transition driven by disorder from the NI phase to the QSH state akin to the emergence of a TAI~\cite{Li2009, Groth2009}. While for 4QLs the intrinsic QSH phase survives at finite disorder first, it turns to a trivial Anderson insulator at larger disorder, which happens for each of the scenarios considered in the following at increased disorder strength (and therefore we will omit this transition in the following discussion \footnote{To discuss this strong disorder phase boundary, a scaling analysis as done in Ref.~\cite{Groth2009} can be performed. This is, however, beyond the scope of this work.}). At non-zero exchange fields, such as $gM=\pm 30$ meV (see Fig. \ref{Mex_disorder} (a) and (c)), the phase diagram is surprisingly rich and the disorder can induce a TPT from a NI first to a QAH phase and then to a spin-Chern insulator in the 3QLs system as the disorder successively inverts the two bands of the block-diagonal Hamiltonian. %Furthermore, QAHI to QSHI transition is also seen at around $M=25$ from (1a)(1b) of Figure \ref{Mex_disorder}.
A similar observation holds for the 4QLs case at large enough ($|gM|> 40$ meV); see Fig.~\ref{Mex_disorder} (b) and (d), when the clean system is in the QAH phase. Here, the disorder induced transition can drive a QAH state to spin-Chern insulator. All of the transitions in the weak disorder region are convincingly reproduced by a self-consistent Born approximation shown as lines in the phase diagrams (see SM \cite{SuppMater}). Although the QAH effect can be realized both in 3QLs and 4QLs TI thin films, their phase boundaries behave differently as disorder strength is increased. In the magnetically doped TI thin films with 3QLs, weak disorder stabilizes the QAH state, which suggests that the critical value of exchange field to induce the QAH effect becomes smaller when we increase the disorder strength. A similar behavior was reported for the quantum Hall effect \cite{PatrickLee1985}. In contrast, for 4QLs thin films, weak disorder stabilizes the spin-Chern insulating phase and a larger exchange field is required to induce the QAH effect compared to the clean sample. These discoveries may help to distinguish the two possible scenarios to achieve the QAH state experimentally \cite{yu2010quantized}.

We further examine the phase diagrams as a function of disorder strength $W$ and tuning the Fermi energy $E_F$ for 3QLs magnetically doped TI thin films (for 4QLs see SM \cite{SuppMater}). The results are summarized in Fig.~\ref{3QL}. At vanishing exchange field, shown in (a) and (b), we find that the disorder decreases the band gap of the thin film, driving it to be metallic. Further enhancing the disorder potential, a topologically non-trivial TAI state emerges from the NI state. Such a TPT is indicated by the conductance $G$ changing from 0 to 2 ($e^2/h$) and the standard deviation $\delta G$ vanishing in these regions (the non-disordered system corresponds to the point D of (e) in Fig.~\ref{Mex_disorder}). The phase boundary curve obtained from the self-consistent Born approximation shows excellent agreement with the full calculations. %When we further increase the disorder potential, the TAI phase will disappear and the whole system becomes a trivial Anderson insulator. 
Our discoveries are consistent with previous discussions about HgTe/CdTe quantum wells \cite{Li2009, Groth2009}.  

%{\color{blue} TO: to describe weak disorder boundary}. {\color{blue} TO: For the discussion of strong disorder phase boundary, the scaling analysis can be performed as it has been done for HgTe/CdTe quantum wells in ~\cite{Groth2009}, which however is beyond the scope of this work.} 

Next, we turn to finite exchange field in panels (c), (d) and (e), (f) with $gM=30$ meV and $gM=52$ meV, whose intrinsic states without disorder belong to the NI and QAH phases, respectively. For the NI state shown in panels (c) and (d), the time reversal symmetry is broken by the long-range FM order, but the exchange field is not strong enough to drive the thin film to be a Chern insulator, it is a trivial dilute magnetic semiconductor. Increasing the disorder strength, the band gap of the thin film becomes smaller in the weak disorder region. Eventually, it closes at the critical value $W_c\approx$ 290 meV and re-opens again, at which point $G$ assumes the quantized value $e^2/h$ and $\delta G$ is zero, indicating the emergence of a QAH state driven by disorder. More interestingly, when we further enhance the disorder potential, the TI thin film will go to the metallic region again and then becomes a spin-Chern insulator ($G=2 e^2/h$ and $\delta G=0$). %When the disorder is strong enough, this thin film will become a topologically trivial Anderson insulator finally. 
The weak disorder part of our observations can again be well understood by the self-consistent Born approximation given as lines (see SM~\cite{SuppMater} for details). %As shown in Eq.~\eqref{born}, 
%The effective mass term in the Hamiltonian (Eq.~\eqref{hamiltonian}) and chemical potential are renormalized by disorder (see SM~\cite{SuppMater}), the change of effective chemical potentials are marked by colored lines in Fig.~\ref{3QL}. 
Increasing the disorder potential first induces a band closing for the lower block Hamiltonian in the weak disorder region ($W_{c, 1} \approx 290$ meV), and then for higher disorder for the upper block Hamiltonian ($W_{c, 2} \approx 670$ meV)). Consequently disorder tunes TI thin films through a series of phase transitions from NI$\to$QAH$\to$spin-Chern insulator. Similar transitions were reported for a less material-oriented toy models before, for example strongly spin-orbit coupled graphene lattices with AFM order \cite{Suying2016} and the Lieb lattice \cite{ChenRui2017}, which, however, have little possibility to be realized from a real materials' point of view. 
%{\color{blue}TO: From the next sentence, maybe it's better to change the paragraph since otherwise this paragraph would be too long and also the story is shifted for discussion of (e)(f)?} 

In Fig.~\ref{3QL} (e) and (f), the exchange field is strong enough such that the system is in a QAH phase at $W=0$ (corresponding to A and G in the sub-panels of Fig.~\ref{Mex_disorder} (e)). %This is because the lower block Hamiltonian turns topologically non-trivial as the system crosses the point marked by F in the sub-panels of Fig.~\ref{Mex_disorder} (e). 
In this case increasing the disorder widens the QAH phase, because the band gap in the lower block Hamiltonian with negative mass is increased by the disorder induced renomormalization (see SM for details~\cite{SuppMater}). For the upper block Hamiltonian whose mass term is positive at $W=0$, the disorder will decrease the band gap. At a critical disorder strength of $W_{c}\approx780$ meV, we observe band inversion, indicating a TPT from the QAH to the spin-Chern insulator phase.
%the gap in the upper block Hamiltonian closes and a transition into a topologically non-trivial band is observed. This drives the system from the QAH to the spin-Chern insulator phase as in the case of panel (c). 
%Both aspects of the gap widening in the lower block Hamiltonian and the gap closing and re-opening in the upper block part 
The band evolution processes in this system are again captured well within the self-consistent Born approximation at small disorder (see SM for details~\cite{SuppMater}). We connect our results directly to the experimental findings of Ref.~\cite{Chang2016Obse}, which reports a phase transition from a QAH to a mixed QAH plus metallic states in V-doped doped (Bi, Sb)$_2$Te$_3$ thin films by changing the gate voltage. Sweeps in the gate voltage correspond to vertical cuts in Fig.~\ref{3QL} and the experimental transition is consistent with the behavior observed around $W=500$ meV in panels (c) and (e) (with corresponding panels (d) and (f)). In our theoretical results starting from $E_F=0$ in the QAH state with finite disorder, one can tune into a regime where $G$ is in between 1 and 2 quantized conductance (yellow color), but the standard deviations $\delta G$ are still quite suppressed (red color), by increasing $E_F$. This is a mixed states with properties in between the QAH and a metallic phase. 

%{\color{blue}TO: for (e)(f), do we directly start discussion with the help of the Born approximation since the sentence "In this case increasing the disorder widens the QAH phase, as the gap in the lower block Hamiltonian......" already use Born analysis? If so, I think we can just remove the final sentence, "Both aspects of the gap widening....." since it would be redundant?} %At larger disorder again a transition to a trivial Anderson insulator is found, which is beyond our weak disorder Born approximation.

%{\color{red}: PZ's comments, Takuya please rewrites Figure3c and Figure4 following the strategy shown in paragraph above, numerical results firstly, then Born Approx, you cannot just use one or two sentence to cover all results. The readers is not as familiar with the results as you can. You should let them know what you are doing first, then explain your simulations.}

%{\color{blue} TO: As an experimental connection, phase transition between QAH and QAH+Metal has been detected in the V-doped doped (Bi, Sb)$_2$Te$_3$ thin film by changing the gate voltage~\cite{Chang2016Obse}, which may correspond to the transition observed along the vertical line of around $W=500$ meV in panel (c)(d)(e)(f).}
%{\color{blue} TO: ($W_{critical} \approx 780$ meV) in panel (e)(f) of Fig \ref{3QL}}

%
%In my opinion we have said this enough of times by now.
%whereas it is not the case for the one to Anderson insulator. 

%Do we really need this again? I don't think so... it could go into a conclusion. summary paragraph if we write one?

\textit{Conclusion}.--- 
We summarize that in all phase diagrams, non-time-reversal-breaking disorder tends to promote the topological character and once the material achieves a QSH or spin-Chern insulator state, for which both bands show a topological character, by increasing disorder, it cannot be driven into another topological state, but only into an Anderson insulator. This observation is consistent with the negative contribution to the renormalization of the topological mass term found from the solution of the non-self consistent Born approximation (see SM~\cite{SuppMater}).  With this, our work puts disordered thin films of magnetically doped (Bi, Sb)$_2$Te$_3$ at the forefront of condensed matter research with a number of interesting TPTs to be confirmed experimentally. Intriguing future avenues of research should address the inclusion of magnetic disorder as well as a self-consistent treatment of the exchange field in the presence of disorder~\cite{work_progress}.%{\color{red} TAKUYA PLS PUT A REFERENCE WITH US AS THE AUTHORS HERE STATING Work in progress. }.

%This is also the case for all phase diagram we show in this work, namely once the system achieves QSH or spin Chern insulator by increasing disorder,

%{\color{blue} TO: ($W_{critical} \approx 210$ meV) in panel (e)(f) of Fig \ref{4QL}}

%\section{Conclusion}

%\noindent \textit{Acknowledgments}---
\begin{acknowledgments} 
\noindent \textit{Acknowledgments}.---
We thank M. Brandbyge
for fruitful discussions.
This work was supported by the Deutsche
Forschungsgemeinschaft (DFG, German Research Foundation) via RTG 1995 and Germany’s Excellence Strategy - Cluster of Excellence Matter and Light for Quantum Computing (ML4Q) EXC 2004/1 - 390534769, by the European Research Council (ERC-2015-AdG-694097), Grupos Consolidados (IT1249-19) and the Flatiron Institute, a division of the Simons Foundation. Simulations were performed with computing resources granted by RWTH Aachen University under project rwth0601 and rwth0507. P.T. acknowledges the support from the Fundamental Research Funds for the Central Universities (ZG216S20A1) and the 111 Project (B17002). We acknowledge support from the Max Planck-New York City Center for Non-Equilibrium Quantum Phenomena. We acknowledge support of the Partner Group of the Max Planck Institute for the Structure and Dynamics of Matter at the School of Materials Science and Engineering, Beihang University, P.R China. 
\end{acknowledgments}

%\bibliography{apssamp}
\bibliography{Manuscript}% 

\providecommand{\noopsort}[1]{}\providecommand{\singleletter}[1]{#1}%
\begin{thebibliography}{64}%
\makeatletter
\providecommand \@ifxundefined [1]{%
 \@ifx{#1\undefined}
}%
\providecommand \@ifnum [1]{%
 \ifnum #1\expandafter \@firstoftwo
 \else \expandafter \@secondoftwo
 \fi
}%
\providecommand \@ifx [1]{%
 \ifx #1\expandafter \@firstoftwo
 \else \expandafter \@secondoftwo
 \fi
}%
\providecommand \natexlab [1]{#1}%
\providecommand \enquote  [1]{``#1''}%
\providecommand \bibnamefont  [1]{#1}%
\providecommand \bibfnamefont [1]{#1}%
\providecommand \citenamefont [1]{#1}%
\providecommand \href@noop [0]{\@secondoftwo}%
\providecommand \href [0]{\begingroup \@sanitize@url \@href}%
\providecommand \@href[1]{\@@startlink{#1}\@@href}%
\providecommand \@@href[1]{\endgroup#1\@@endlink}%
\providecommand \@sanitize@url [0]{\catcode `\\12\catcode `\$12\catcode
  `\&12\catcode `\#12\catcode `\^12\catcode `\_12\catcode `\%12\relax}%
\providecommand \@@startlink[1]{}%
\providecommand \@@endlink[0]{}%
\providecommand \url  [0]{\begingroup\@sanitize@url \@url }%
\providecommand \@url [1]{\endgroup\@href {#1}{\urlprefix }}%
\providecommand \urlprefix  [0]{URL }%
\providecommand \Eprint [0]{\href }%
\providecommand \doibase [0]{http://dx.doi.org/}%
\providecommand \selectlanguage [0]{\@gobble}%
\providecommand \bibinfo  [0]{\@secondoftwo}%
\providecommand \bibfield  [0]{\@secondoftwo}%
\providecommand \translation [1]{[#1]}%
\providecommand \BibitemOpen [0]{}%
\providecommand \bibitemStop [0]{}%
\providecommand \bibitemNoStop [0]{.\EOS\space}%
\providecommand \EOS [0]{\spacefactor3000\relax}%
\providecommand \BibitemShut  [1]{\csname bibitem#1\endcsname}%
\let\auto@bib@innerbib\@empty
%</preamble>
\bibitem [{\citenamefont {Mong}\ \emph {et~al.}(2010)\citenamefont {Mong},
  \citenamefont {Essin},\ and\ \citenamefont {Moore}}]{Mong2010}%
  \BibitemOpen
  \bibfield  {author} {\bibinfo {author} {\bibfnamefont {R.~S.~K.}\
  \bibnamefont {Mong}}, \bibinfo {author} {\bibfnamefont {A.~M.}\ \bibnamefont
  {Essin}}, \ and\ \bibinfo {author} {\bibfnamefont {J.~E.}\ \bibnamefont
  {Moore}},\ }\href {\doibase 10.1103/PhysRevB.81.245209} {\bibfield  {journal}
  {\bibinfo  {journal} {Phys. Rev. B}\ }\textbf {\bibinfo {volume} {81}},\
  \bibinfo {pages} {245209} (\bibinfo {year} {2010})}\BibitemShut {NoStop}%
\bibitem [{\citenamefont {Otrokov}\ \emph {et~al.}(2019)\citenamefont
  {Otrokov}, \citenamefont {Klimovskikh}, \citenamefont {Bentmann},
  \citenamefont {Estyunin}, \citenamefont {Zeugner}, \citenamefont {Aliev},
  \citenamefont {Ga{\ss}}, \citenamefont {Wolter}, \citenamefont {Koroleva},
  \citenamefont {Shikin} \emph {et~al.}}]{otrokov2019prediction}%
  \BibitemOpen
  \bibfield  {author} {\bibinfo {author} {\bibfnamefont {M.~M.}\ \bibnamefont
  {Otrokov}}, \bibinfo {author} {\bibfnamefont {I.~I.}\ \bibnamefont
  {Klimovskikh}}, \bibinfo {author} {\bibfnamefont {H.}~\bibnamefont
  {Bentmann}}, \bibinfo {author} {\bibfnamefont {D.}~\bibnamefont {Estyunin}},
  \bibinfo {author} {\bibfnamefont {A.}~\bibnamefont {Zeugner}}, \bibinfo
  {author} {\bibfnamefont {Z.~S.}\ \bibnamefont {Aliev}}, \bibinfo {author}
  {\bibfnamefont {S.}~\bibnamefont {Ga{\ss}}}, \bibinfo {author} {\bibfnamefont
  {A.}~\bibnamefont {Wolter}}, \bibinfo {author} {\bibfnamefont
  {A.}~\bibnamefont {Koroleva}}, \bibinfo {author} {\bibfnamefont {A.~M.}\
  \bibnamefont {Shikin}},  \emph {et~al.},\ }\href {\doibase
  10.1038/s41586-019-1840-9} {\bibfield  {journal} {\bibinfo  {journal}
  {Nature}\ }\textbf {\bibinfo {volume} {576}},\ \bibinfo {pages} {416}
  (\bibinfo {year} {2019})}\BibitemShut {NoStop}%
\bibitem [{\citenamefont {Li}\ \emph {et~al.}(2019)\citenamefont {Li},
  \citenamefont {Li}, \citenamefont {Du}, \citenamefont {Wang}, \citenamefont
  {Gu}, \citenamefont {Zhang}, \citenamefont {He}, \citenamefont {Duan},\ and\
  \citenamefont {Xu}}]{liJH2018}%
  \BibitemOpen
  \bibfield  {author} {\bibinfo {author} {\bibfnamefont {J.}~\bibnamefont
  {Li}}, \bibinfo {author} {\bibfnamefont {Y.}~\bibnamefont {Li}}, \bibinfo
  {author} {\bibfnamefont {S.}~\bibnamefont {Du}}, \bibinfo {author}
  {\bibfnamefont {Z.}~\bibnamefont {Wang}}, \bibinfo {author} {\bibfnamefont
  {B.-L.}\ \bibnamefont {Gu}}, \bibinfo {author} {\bibfnamefont {S.-C.}\
  \bibnamefont {Zhang}}, \bibinfo {author} {\bibfnamefont {K.}~\bibnamefont
  {He}}, \bibinfo {author} {\bibfnamefont {W.}~\bibnamefont {Duan}}, \ and\
  \bibinfo {author} {\bibfnamefont {Y.}~\bibnamefont {Xu}},\ }\href@noop {}
  {\bibfield  {journal} {\bibinfo  {journal} {Sci. Adv.}\ }\textbf {\bibinfo
  {volume} {\textbf{5}}},\ \bibinfo {pages} {eaaw5685} (\bibinfo {year}
  {2019})}\BibitemShut {NoStop}%
\bibitem [{\citenamefont {Tang}\ \emph {et~al.}(2016)\citenamefont {Tang},
  \citenamefont {Zhou}, \citenamefont {Xu},\ and\ \citenamefont
  {Zhang}}]{tang2016dirac}%
  \BibitemOpen
  \bibfield  {author} {\bibinfo {author} {\bibfnamefont {P.}~\bibnamefont
  {Tang}}, \bibinfo {author} {\bibfnamefont {Q.}~\bibnamefont {Zhou}}, \bibinfo
  {author} {\bibfnamefont {G.}~\bibnamefont {Xu}}, \ and\ \bibinfo {author}
  {\bibfnamefont {S.-C.}\ \bibnamefont {Zhang}},\ }\href@noop {} {\bibfield
  {journal} {\bibinfo  {journal} {Nat. Phys.}\ }\textbf {\bibinfo {volume}
  {\textbf{12}}},\ \bibinfo {pages} {1100} (\bibinfo {year}
  {2016})}\BibitemShut {NoStop}%
\bibitem [{\citenamefont {Armitage}\ \emph {et~al.}(2018)\citenamefont
  {Armitage}, \citenamefont {Mele},\ and\ \citenamefont
  {Vishwanath}}]{Armitage2018}%
  \BibitemOpen
  \bibfield  {author} {\bibinfo {author} {\bibfnamefont {N.~P.}\ \bibnamefont
  {Armitage}}, \bibinfo {author} {\bibfnamefont {E.~J.}\ \bibnamefont {Mele}},
  \ and\ \bibinfo {author} {\bibfnamefont {A.}~\bibnamefont {Vishwanath}},\
  }\href {\doibase 10.1103/RevModPhys.90.015001} {\bibfield  {journal}
  {\bibinfo  {journal} {Rev. Mod. Phys.}\ }\textbf {\bibinfo {volume} {90}},\
  \bibinfo {pages} {015001} (\bibinfo {year} {2018})}\BibitemShut {NoStop}%
\bibitem [{\citenamefont {Yu}\ \emph {et~al.}(2010)\citenamefont {Yu},
  \citenamefont {Zhang}, \citenamefont {Zhang}, \citenamefont {Zhang},
  \citenamefont {Dai},\ and\ \citenamefont {Fang}}]{yu2010quantized}%
  \BibitemOpen
  \bibfield  {author} {\bibinfo {author} {\bibfnamefont {R.}~\bibnamefont
  {Yu}}, \bibinfo {author} {\bibfnamefont {W.}~\bibnamefont {Zhang}}, \bibinfo
  {author} {\bibfnamefont {H.-J.}\ \bibnamefont {Zhang}}, \bibinfo {author}
  {\bibfnamefont {S.-C.}\ \bibnamefont {Zhang}}, \bibinfo {author}
  {\bibfnamefont {X.}~\bibnamefont {Dai}}, \ and\ \bibinfo {author}
  {\bibfnamefont {Z.}~\bibnamefont {Fang}},\ }\href {\doibase
  10.1126/science.1187485} {\bibfield  {journal} {\bibinfo  {journal}
  {Science}\ }\textbf {\bibinfo {volume} {329}},\ \bibinfo {pages} {61}
  (\bibinfo {year} {2010})}\BibitemShut {NoStop}%
\bibitem [{\citenamefont {Chang}\ \emph
  {et~al.}(2013{\natexlab{a}})\citenamefont {Chang}, \citenamefont {Zhang},
  \citenamefont {Feng}, \citenamefont {Shen}, \citenamefont {Zhang},
  \citenamefont {Guo}, \citenamefont {Li}, \citenamefont {Ou}, \citenamefont
  {Wei}, \citenamefont {Wang} \emph {et~al.}}]{chang2013experimental}%
  \BibitemOpen
  \bibfield  {author} {\bibinfo {author} {\bibfnamefont {C.-Z.}\ \bibnamefont
  {Chang}}, \bibinfo {author} {\bibfnamefont {J.}~\bibnamefont {Zhang}},
  \bibinfo {author} {\bibfnamefont {X.}~\bibnamefont {Feng}}, \bibinfo {author}
  {\bibfnamefont {J.}~\bibnamefont {Shen}}, \bibinfo {author} {\bibfnamefont
  {Z.}~\bibnamefont {Zhang}}, \bibinfo {author} {\bibfnamefont
  {M.}~\bibnamefont {Guo}}, \bibinfo {author} {\bibfnamefont {K.}~\bibnamefont
  {Li}}, \bibinfo {author} {\bibfnamefont {Y.}~\bibnamefont {Ou}}, \bibinfo
  {author} {\bibfnamefont {P.}~\bibnamefont {Wei}}, \bibinfo {author}
  {\bibfnamefont {L.-L.}\ \bibnamefont {Wang}},  \emph {et~al.},\ }\href
  {\doibase 10.1126/science.1234414} {\bibfield  {journal} {\bibinfo  {journal}
  {Science}\ }\textbf {\bibinfo {volume} {340}},\ \bibinfo {pages} {167}
  (\bibinfo {year} {2013}{\natexlab{a}})}\BibitemShut {NoStop}%
\bibitem [{\citenamefont {Liu}\ \emph {et~al.}(2016)\citenamefont {Liu},
  \citenamefont {Zhang},\ and\ \citenamefont {Qi}}]{liu2016quantum}%
  \BibitemOpen
  \bibfield  {author} {\bibinfo {author} {\bibfnamefont {C.-X.}\ \bibnamefont
  {Liu}}, \bibinfo {author} {\bibfnamefont {S.-C.}\ \bibnamefont {Zhang}}, \
  and\ \bibinfo {author} {\bibfnamefont {X.-L.}\ \bibnamefont {Qi}},\ }\href
  {\doibase 10.1146/annurev-conmatphys-031115-011417} {\bibfield  {journal}
  {\bibinfo  {journal} {Annu. Rev. Condens. Matter Phys.}\ }\textbf {\bibinfo
  {volume} {7}},\ \bibinfo {pages} {301} (\bibinfo {year} {2016})}\BibitemShut
  {NoStop}%
\bibitem [{\citenamefont {He}\ \emph {et~al.}(2018)\citenamefont {He},
  \citenamefont {Wang},\ and\ \citenamefont {Xue}}]{he2018topological}%
  \BibitemOpen
  \bibfield  {author} {\bibinfo {author} {\bibfnamefont {K.}~\bibnamefont
  {He}}, \bibinfo {author} {\bibfnamefont {Y.}~\bibnamefont {Wang}}, \ and\
  \bibinfo {author} {\bibfnamefont {Q.-K.}\ \bibnamefont {Xue}},\ }\href
  {\doibase 10.1146/annurev-conmatphys-033117-054144} {\bibfield  {journal}
  {\bibinfo  {journal} {Annu. Rev. Condens. Matter Phys.}\ }\textbf {\bibinfo
  {volume} {9}},\ \bibinfo {pages} {329} (\bibinfo {year} {2018})}\BibitemShut
  {NoStop}%
\bibitem [{\citenamefont {Chang}\ \emph
  {et~al.}(2013{\natexlab{b}})\citenamefont {Chang}, \citenamefont {Zhang},
  \citenamefont {Liu}, \citenamefont {Zhang}, \citenamefont {Feng},
  \citenamefont {Li}, \citenamefont {Wang}, \citenamefont {Chen}, \citenamefont
  {Dai}, \citenamefont {Fang}, \citenamefont {Qi}, \citenamefont {Zhang},
  \citenamefont {Wang}, \citenamefont {He}, \citenamefont {Ma},\ and\
  \citenamefont {Xue}}]{ChangCZ2013}%
  \BibitemOpen
  \bibfield  {author} {\bibinfo {author} {\bibfnamefont {C.-Z.}\ \bibnamefont
  {Chang}}, \bibinfo {author} {\bibfnamefont {J.}~\bibnamefont {Zhang}},
  \bibinfo {author} {\bibfnamefont {M.}~\bibnamefont {Liu}}, \bibinfo {author}
  {\bibfnamefont {Z.}~\bibnamefont {Zhang}}, \bibinfo {author} {\bibfnamefont
  {X.}~\bibnamefont {Feng}}, \bibinfo {author} {\bibfnamefont {K.}~\bibnamefont
  {Li}}, \bibinfo {author} {\bibfnamefont {L.-L.}\ \bibnamefont {Wang}},
  \bibinfo {author} {\bibfnamefont {X.}~\bibnamefont {Chen}}, \bibinfo {author}
  {\bibfnamefont {X.}~\bibnamefont {Dai}}, \bibinfo {author} {\bibfnamefont
  {Z.}~\bibnamefont {Fang}}, \bibinfo {author} {\bibfnamefont {X.-L.}\
  \bibnamefont {Qi}}, \bibinfo {author} {\bibfnamefont {S.-C.}\ \bibnamefont
  {Zhang}}, \bibinfo {author} {\bibfnamefont {Y.}~\bibnamefont {Wang}},
  \bibinfo {author} {\bibfnamefont {K.}~\bibnamefont {He}}, \bibinfo {author}
  {\bibfnamefont {X.-C.}\ \bibnamefont {Ma}}, \ and\ \bibinfo {author}
  {\bibfnamefont {Q.-K.}\ \bibnamefont {Xue}},\ }\href {\doibase
  10.1002/adma.201203493} {\bibfield  {journal} {\bibinfo  {journal} {Adv.
  Mater.}\ }\textbf {\bibinfo {volume} {25}},\ \bibinfo {pages} {1065}
  (\bibinfo {year} {2013}{\natexlab{b}})}\BibitemShut {NoStop}%
\bibitem [{\citenamefont {Zhang}\ \emph {et~al.}(2013)\citenamefont {Zhang},
  \citenamefont {Chang}, \citenamefont {Tang}, \citenamefont {Zhang},
  \citenamefont {Feng}, \citenamefont {Li}, \citenamefont {Wang}, \citenamefont
  {Chen}, \citenamefont {Liu}, \citenamefont {Duan} \emph
  {et~al.}}]{zhang2013topology}%
  \BibitemOpen
  \bibfield  {author} {\bibinfo {author} {\bibfnamefont {J.}~\bibnamefont
  {Zhang}}, \bibinfo {author} {\bibfnamefont {C.-Z.}\ \bibnamefont {Chang}},
  \bibinfo {author} {\bibfnamefont {P.}~\bibnamefont {Tang}}, \bibinfo {author}
  {\bibfnamefont {Z.}~\bibnamefont {Zhang}}, \bibinfo {author} {\bibfnamefont
  {X.}~\bibnamefont {Feng}}, \bibinfo {author} {\bibfnamefont {K.}~\bibnamefont
  {Li}}, \bibinfo {author} {\bibfnamefont {L.-l.}\ \bibnamefont {Wang}},
  \bibinfo {author} {\bibfnamefont {X.}~\bibnamefont {Chen}}, \bibinfo {author}
  {\bibfnamefont {C.}~\bibnamefont {Liu}}, \bibinfo {author} {\bibfnamefont
  {W.}~\bibnamefont {Duan}},  \emph {et~al.},\ }\href {\doibase
  10.1126/science.1230905} {\bibfield  {journal} {\bibinfo  {journal}
  {Science}\ }\textbf {\bibinfo {volume} {339}},\ \bibinfo {pages} {1582}
  (\bibinfo {year} {2013})}\BibitemShut {NoStop}%
\bibitem [{\citenamefont {Sharpe}\ \emph {et~al.}(2019)\citenamefont {Sharpe},
  \citenamefont {Fox}, \citenamefont {Barnard}, \citenamefont {Finney},
  \citenamefont {Watanabe}, \citenamefont {Taniguchi}, \citenamefont
  {Kastner},\ and\ \citenamefont {Goldhaber-Gordon}}]{sharpe2019emergent}%
  \BibitemOpen
  \bibfield  {author} {\bibinfo {author} {\bibfnamefont {A.~L.}\ \bibnamefont
  {Sharpe}}, \bibinfo {author} {\bibfnamefont {E.~J.}\ \bibnamefont {Fox}},
  \bibinfo {author} {\bibfnamefont {A.~W.}\ \bibnamefont {Barnard}}, \bibinfo
  {author} {\bibfnamefont {J.}~\bibnamefont {Finney}}, \bibinfo {author}
  {\bibfnamefont {K.}~\bibnamefont {Watanabe}}, \bibinfo {author}
  {\bibfnamefont {T.}~\bibnamefont {Taniguchi}}, \bibinfo {author}
  {\bibfnamefont {M.}~\bibnamefont {Kastner}}, \ and\ \bibinfo {author}
  {\bibfnamefont {D.}~\bibnamefont {Goldhaber-Gordon}},\ }\href {\doibase
  10.1126/science.aaw3780} {\bibfield  {journal} {\bibinfo  {journal}
  {Science}\ }\textbf {\bibinfo {volume} {365}},\ \bibinfo {pages} {605}
  (\bibinfo {year} {2019})}\BibitemShut {NoStop}%
\bibitem [{\citenamefont {Serlin}\ \emph {et~al.}(2020)\citenamefont {Serlin},
  \citenamefont {Tschirhart}, \citenamefont {Polshyn}, \citenamefont {Zhang},
  \citenamefont {Zhu}, \citenamefont {Watanabe}, \citenamefont {Taniguchi},
  \citenamefont {Balents},\ and\ \citenamefont {Young}}]{serlin2020intrinsic}%
  \BibitemOpen
  \bibfield  {author} {\bibinfo {author} {\bibfnamefont {M.}~\bibnamefont
  {Serlin}}, \bibinfo {author} {\bibfnamefont {C.}~\bibnamefont {Tschirhart}},
  \bibinfo {author} {\bibfnamefont {H.}~\bibnamefont {Polshyn}}, \bibinfo
  {author} {\bibfnamefont {Y.}~\bibnamefont {Zhang}}, \bibinfo {author}
  {\bibfnamefont {J.}~\bibnamefont {Zhu}}, \bibinfo {author} {\bibfnamefont
  {K.}~\bibnamefont {Watanabe}}, \bibinfo {author} {\bibfnamefont
  {T.}~\bibnamefont {Taniguchi}}, \bibinfo {author} {\bibfnamefont
  {L.}~\bibnamefont {Balents}}, \ and\ \bibinfo {author} {\bibfnamefont
  {A.}~\bibnamefont {Young}},\ }\href {\doibase 10.1126/science.aay5533}
  {\bibfield  {journal} {\bibinfo  {journal} {Science}\ }\textbf {\bibinfo
  {volume} {367}},\ \bibinfo {pages} {900} (\bibinfo {year}
  {2020})}\BibitemShut {NoStop}%
\bibitem [{\citenamefont {Deng}\ \emph {et~al.}(2020)\citenamefont {Deng},
  \citenamefont {Yu}, \citenamefont {Shi}, \citenamefont {Guo}, \citenamefont
  {Xu}, \citenamefont {Wang}, \citenamefont {Chen},\ and\ \citenamefont
  {Zhang}}]{deng2020quantum}%
  \BibitemOpen
  \bibfield  {author} {\bibinfo {author} {\bibfnamefont {Y.}~\bibnamefont
  {Deng}}, \bibinfo {author} {\bibfnamefont {Y.}~\bibnamefont {Yu}}, \bibinfo
  {author} {\bibfnamefont {M.~Z.}\ \bibnamefont {Shi}}, \bibinfo {author}
  {\bibfnamefont {Z.}~\bibnamefont {Guo}}, \bibinfo {author} {\bibfnamefont
  {Z.}~\bibnamefont {Xu}}, \bibinfo {author} {\bibfnamefont {J.}~\bibnamefont
  {Wang}}, \bibinfo {author} {\bibfnamefont {X.~H.}\ \bibnamefont {Chen}}, \
  and\ \bibinfo {author} {\bibfnamefont {Y.}~\bibnamefont {Zhang}},\ }\href
  {\doibase 10.1126/science.aax8156} {\bibfield  {journal} {\bibinfo  {journal}
  {Science}\ }\textbf {\bibinfo {volume} {367}},\ \bibinfo {pages} {895}
  (\bibinfo {year} {2020})}\BibitemShut {NoStop}%
\bibitem [{\citenamefont {Checkelsky}\ \emph {et~al.}(2014)\citenamefont
  {Checkelsky}, \citenamefont {Yoshimi}, \citenamefont {Tsukazaki},
  \citenamefont {Takahashi}, \citenamefont {Kozuka}, \citenamefont {Falson},
  \citenamefont {Kawasaki},\ and\ \citenamefont
  {Tokura}}]{Checkelsky2014Trajectory}%
  \BibitemOpen
  \bibfield  {author} {\bibinfo {author} {\bibfnamefont {J.~G.}\ \bibnamefont
  {Checkelsky}}, \bibinfo {author} {\bibfnamefont {R.}~\bibnamefont {Yoshimi}},
  \bibinfo {author} {\bibfnamefont {A.}~\bibnamefont {Tsukazaki}}, \bibinfo
  {author} {\bibfnamefont {K.~S.}\ \bibnamefont {Takahashi}}, \bibinfo {author}
  {\bibfnamefont {Y.}~\bibnamefont {Kozuka}}, \bibinfo {author} {\bibfnamefont
  {J.}~\bibnamefont {Falson}}, \bibinfo {author} {\bibfnamefont
  {M.}~\bibnamefont {Kawasaki}}, \ and\ \bibinfo {author} {\bibfnamefont
  {Y.}~\bibnamefont {Tokura}},\ }\href {\doibase 10.1038/nphys3053} {\bibfield
  {journal} {\bibinfo  {journal} {Nat. Phys.}\ }\textbf {\bibinfo {volume}
  {10}},\ \bibinfo {pages} {731} (\bibinfo {year} {2014})}\BibitemShut
  {NoStop}%
\bibitem [{\citenamefont {Chang}\ \emph {et~al.}(2015)\citenamefont {Chang},
  \citenamefont {Zhao}, \citenamefont {Kim}, \citenamefont {Zhang},
  \citenamefont {Assaf}, \citenamefont {Heiman}, \citenamefont {Zhang},
  \citenamefont {Liu}, \citenamefont {Chan},\ and\ \citenamefont
  {Moodera}}]{chang2015high}%
  \BibitemOpen
  \bibfield  {author} {\bibinfo {author} {\bibfnamefont {C.-Z.}\ \bibnamefont
  {Chang}}, \bibinfo {author} {\bibfnamefont {W.}~\bibnamefont {Zhao}},
  \bibinfo {author} {\bibfnamefont {D.~Y.}\ \bibnamefont {Kim}}, \bibinfo
  {author} {\bibfnamefont {H.}~\bibnamefont {Zhang}}, \bibinfo {author}
  {\bibfnamefont {B.~A.}\ \bibnamefont {Assaf}}, \bibinfo {author}
  {\bibfnamefont {D.}~\bibnamefont {Heiman}}, \bibinfo {author} {\bibfnamefont
  {S.-C.}\ \bibnamefont {Zhang}}, \bibinfo {author} {\bibfnamefont
  {C.}~\bibnamefont {Liu}}, \bibinfo {author} {\bibfnamefont {M.~H.}\
  \bibnamefont {Chan}}, \ and\ \bibinfo {author} {\bibfnamefont {J.~S.}\
  \bibnamefont {Moodera}},\ }\href {\doibase 10.1038/nmat4204} {\bibfield
  {journal} {\bibinfo  {journal} {Nat. Mater.}\ }\textbf {\bibinfo {volume}
  {14}},\ \bibinfo {pages} {473} (\bibinfo {year} {2015})}\BibitemShut
  {NoStop}%
\bibitem [{\citenamefont {Feng}\ \emph {et~al.}(2015)\citenamefont {Feng},
  \citenamefont {Feng}, \citenamefont {Ou}, \citenamefont {Wang}, \citenamefont
  {Liu}, \citenamefont {Zhang}, \citenamefont {Zhao}, \citenamefont {Jiang},
  \citenamefont {Zhang}, \citenamefont {He}, \citenamefont {Ma}, \citenamefont
  {Xue},\ and\ \citenamefont {Wang}}]{FengYang2015}%
  \BibitemOpen
  \bibfield  {author} {\bibinfo {author} {\bibfnamefont {Y.}~\bibnamefont
  {Feng}}, \bibinfo {author} {\bibfnamefont {X.}~\bibnamefont {Feng}}, \bibinfo
  {author} {\bibfnamefont {Y.}~\bibnamefont {Ou}}, \bibinfo {author}
  {\bibfnamefont {J.}~\bibnamefont {Wang}}, \bibinfo {author} {\bibfnamefont
  {C.}~\bibnamefont {Liu}}, \bibinfo {author} {\bibfnamefont {L.}~\bibnamefont
  {Zhang}}, \bibinfo {author} {\bibfnamefont {D.}~\bibnamefont {Zhao}},
  \bibinfo {author} {\bibfnamefont {G.}~\bibnamefont {Jiang}}, \bibinfo
  {author} {\bibfnamefont {S.-C.}\ \bibnamefont {Zhang}}, \bibinfo {author}
  {\bibfnamefont {K.}~\bibnamefont {He}}, \bibinfo {author} {\bibfnamefont
  {X.}~\bibnamefont {Ma}}, \bibinfo {author} {\bibfnamefont {Q.-K.}\
  \bibnamefont {Xue}}, \ and\ \bibinfo {author} {\bibfnamefont
  {Y.}~\bibnamefont {Wang}},\ }\href {\doibase 10.1103/PhysRevLett.115.126801}
  {\bibfield  {journal} {\bibinfo  {journal} {Phys. Rev. Lett.}\ }\textbf
  {\bibinfo {volume} {115}},\ \bibinfo {pages} {126801} (\bibinfo {year}
  {2015})}\BibitemShut {NoStop}%
\bibitem [{\citenamefont {Kou}\ \emph {et~al.}(2014)\citenamefont {Kou},
  \citenamefont {Guo}, \citenamefont {Fan}, \citenamefont {Pan}, \citenamefont
  {Lang}, \citenamefont {Jiang}, \citenamefont {Shao}, \citenamefont {Nie},
  \citenamefont {Murata}, \citenamefont {Tang}, \citenamefont {Wang},
  \citenamefont {He}, \citenamefont {Lee}, \citenamefont {Lee},\ and\
  \citenamefont {Wang}}]{Kou2014}%
  \BibitemOpen
  \bibfield  {author} {\bibinfo {author} {\bibfnamefont {X.}~\bibnamefont
  {Kou}}, \bibinfo {author} {\bibfnamefont {S.-T.}\ \bibnamefont {Guo}},
  \bibinfo {author} {\bibfnamefont {Y.}~\bibnamefont {Fan}}, \bibinfo {author}
  {\bibfnamefont {L.}~\bibnamefont {Pan}}, \bibinfo {author} {\bibfnamefont
  {M.}~\bibnamefont {Lang}}, \bibinfo {author} {\bibfnamefont {Y.}~\bibnamefont
  {Jiang}}, \bibinfo {author} {\bibfnamefont {Q.}~\bibnamefont {Shao}},
  \bibinfo {author} {\bibfnamefont {T.}~\bibnamefont {Nie}}, \bibinfo {author}
  {\bibfnamefont {K.}~\bibnamefont {Murata}}, \bibinfo {author} {\bibfnamefont
  {J.}~\bibnamefont {Tang}}, \bibinfo {author} {\bibfnamefont {Y.}~\bibnamefont
  {Wang}}, \bibinfo {author} {\bibfnamefont {L.}~\bibnamefont {He}}, \bibinfo
  {author} {\bibfnamefont {T.-K.}\ \bibnamefont {Lee}}, \bibinfo {author}
  {\bibfnamefont {W.-L.}\ \bibnamefont {Lee}}, \ and\ \bibinfo {author}
  {\bibfnamefont {K.~L.}\ \bibnamefont {Wang}},\ }\href {\doibase
  10.1103/PhysRevLett.113.137201} {\bibfield  {journal} {\bibinfo  {journal}
  {Phys. Rev. Lett.}\ }\textbf {\bibinfo {volume} {113}},\ \bibinfo {pages}
  {137201} (\bibinfo {year} {2014})}\BibitemShut {NoStop}%
\bibitem [{\citenamefont {Bestwick}\ \emph {et~al.}(2015)\citenamefont
  {Bestwick}, \citenamefont {Fox}, \citenamefont {Kou}, \citenamefont {Pan},
  \citenamefont {Wang},\ and\ \citenamefont {Goldhaber-Gordon}}]{Bestwick2015}%
  \BibitemOpen
  \bibfield  {author} {\bibinfo {author} {\bibfnamefont {A.~J.}\ \bibnamefont
  {Bestwick}}, \bibinfo {author} {\bibfnamefont {E.~J.}\ \bibnamefont {Fox}},
  \bibinfo {author} {\bibfnamefont {X.}~\bibnamefont {Kou}}, \bibinfo {author}
  {\bibfnamefont {L.}~\bibnamefont {Pan}}, \bibinfo {author} {\bibfnamefont
  {K.~L.}\ \bibnamefont {Wang}}, \ and\ \bibinfo {author} {\bibfnamefont
  {D.}~\bibnamefont {Goldhaber-Gordon}},\ }\href {\doibase
  10.1103/PhysRevLett.114.187201} {\bibfield  {journal} {\bibinfo  {journal}
  {Phys. Rev. Lett.}\ }\textbf {\bibinfo {volume} {114}},\ \bibinfo {pages}
  {187201} (\bibinfo {year} {2015})}\BibitemShut {NoStop}%
\bibitem [{\citenamefont {Chang}\ \emph {et~al.}(2016)\citenamefont {Chang},
  \citenamefont {Zhao}, \citenamefont {Li}, \citenamefont {Jain}, \citenamefont
  {Liu}, \citenamefont {Moodera},\ and\ \citenamefont {Chan}}]{Chang2016Obse}%
  \BibitemOpen
  \bibfield  {author} {\bibinfo {author} {\bibfnamefont {C.-Z.}\ \bibnamefont
  {Chang}}, \bibinfo {author} {\bibfnamefont {W.}~\bibnamefont {Zhao}},
  \bibinfo {author} {\bibfnamefont {J.}~\bibnamefont {Li}}, \bibinfo {author}
  {\bibfnamefont {J.~K.}\ \bibnamefont {Jain}}, \bibinfo {author}
  {\bibfnamefont {C.}~\bibnamefont {Liu}}, \bibinfo {author} {\bibfnamefont
  {J.~S.}\ \bibnamefont {Moodera}}, \ and\ \bibinfo {author} {\bibfnamefont
  {M.~H.~W.}\ \bibnamefont {Chan}},\ }\href {\doibase
  10.1103/PhysRevLett.117.126802} {\bibfield  {journal} {\bibinfo  {journal}
  {Phys. Rev. Lett.}\ }\textbf {\bibinfo {volume} {117}},\ \bibinfo {pages}
  {126802} (\bibinfo {year} {2016})}\BibitemShut {NoStop}%
\bibitem [{\citenamefont {Yasuda}\ \emph {et~al.}(2017)\citenamefont {Yasuda},
  \citenamefont {Mogi}, \citenamefont {Yoshimi}, \citenamefont {Tsukazaki},
  \citenamefont {Takahashi}, \citenamefont {Kawasaki}, \citenamefont {Kagawa},\
  and\ \citenamefont {Tokura}}]{Yasuda2017Quantized}%
  \BibitemOpen
  \bibfield  {author} {\bibinfo {author} {\bibfnamefont {K.}~\bibnamefont
  {Yasuda}}, \bibinfo {author} {\bibfnamefont {M.}~\bibnamefont {Mogi}},
  \bibinfo {author} {\bibfnamefont {R.}~\bibnamefont {Yoshimi}}, \bibinfo
  {author} {\bibfnamefont {A.}~\bibnamefont {Tsukazaki}}, \bibinfo {author}
  {\bibfnamefont {K.~S.}\ \bibnamefont {Takahashi}}, \bibinfo {author}
  {\bibfnamefont {M.}~\bibnamefont {Kawasaki}}, \bibinfo {author}
  {\bibfnamefont {F.}~\bibnamefont {Kagawa}}, \ and\ \bibinfo {author}
  {\bibfnamefont {Y.}~\bibnamefont {Tokura}},\ }\href {\doibase
  10.1126/science.aan5991} {\bibfield  {journal} {\bibinfo  {journal}
  {Science}\ }\textbf {\bibinfo {volume} {358}},\ \bibinfo {pages} {1311}
  (\bibinfo {year} {2017})}\BibitemShut {NoStop}%
\bibitem [{\citenamefont {Mogi}\ \emph {et~al.}(2015)\citenamefont {Mogi},
  \citenamefont {Yoshimi}, \citenamefont {Tsukazaki}, \citenamefont {Yasuda},
  \citenamefont {Kozuka}, \citenamefont {Takahashi}, \citenamefont {Kawasaki},\
  and\ \citenamefont {Tokura}}]{mogi2015magnetic}%
  \BibitemOpen
  \bibfield  {author} {\bibinfo {author} {\bibfnamefont {M.}~\bibnamefont
  {Mogi}}, \bibinfo {author} {\bibfnamefont {R.}~\bibnamefont {Yoshimi}},
  \bibinfo {author} {\bibfnamefont {A.}~\bibnamefont {Tsukazaki}}, \bibinfo
  {author} {\bibfnamefont {K.}~\bibnamefont {Yasuda}}, \bibinfo {author}
  {\bibfnamefont {Y.}~\bibnamefont {Kozuka}}, \bibinfo {author} {\bibfnamefont
  {K.}~\bibnamefont {Takahashi}}, \bibinfo {author} {\bibfnamefont
  {M.}~\bibnamefont {Kawasaki}}, \ and\ \bibinfo {author} {\bibfnamefont
  {Y.}~\bibnamefont {Tokura}},\ }\href {\doibase doi.org/10.1063/1.4935075}
  {\bibfield  {journal} {\bibinfo  {journal} {Appl. Phys. Lett.}\ }\textbf
  {\bibinfo {volume} {107}},\ \bibinfo {pages} {182401} (\bibinfo {year}
  {2015})}\BibitemShut {NoStop}%
\bibitem [{\citenamefont {Liu}\ \emph {et~al.}(2009)\citenamefont {Liu},
  \citenamefont {Liu}, \citenamefont {Xu}, \citenamefont {Qi},\ and\
  \citenamefont {Zhang}}]{LiuQ2009}%
  \BibitemOpen
  \bibfield  {author} {\bibinfo {author} {\bibfnamefont {Q.}~\bibnamefont
  {Liu}}, \bibinfo {author} {\bibfnamefont {C.-X.}\ \bibnamefont {Liu}},
  \bibinfo {author} {\bibfnamefont {C.}~\bibnamefont {Xu}}, \bibinfo {author}
  {\bibfnamefont {X.-L.}\ \bibnamefont {Qi}}, \ and\ \bibinfo {author}
  {\bibfnamefont {S.-C.}\ \bibnamefont {Zhang}},\ }\href {\doibase
  10.1103/PhysRevLett.102.156603} {\bibfield  {journal} {\bibinfo  {journal}
  {Phys. Rev. Lett.}\ }\textbf {\bibinfo {volume} {102}},\ \bibinfo {pages}
  {156603} (\bibinfo {year} {2009})}\BibitemShut {NoStop}%
\bibitem [{\citenamefont {Checkelsky}\ \emph {et~al.}(2012)\citenamefont
  {Checkelsky}, \citenamefont {Ye}, \citenamefont {Onose}, \citenamefont
  {Iwasa},\ and\ \citenamefont {Tokura}}]{checkelsky2012dirac}%
  \BibitemOpen
  \bibfield  {author} {\bibinfo {author} {\bibfnamefont {J.~G.}\ \bibnamefont
  {Checkelsky}}, \bibinfo {author} {\bibfnamefont {J.}~\bibnamefont {Ye}},
  \bibinfo {author} {\bibfnamefont {Y.}~\bibnamefont {Onose}}, \bibinfo
  {author} {\bibfnamefont {Y.}~\bibnamefont {Iwasa}}, \ and\ \bibinfo {author}
  {\bibfnamefont {Y.}~\bibnamefont {Tokura}},\ }\href {\doibase
  10.1038/nphys2388} {\bibfield  {journal} {\bibinfo  {journal} {Nat. Phys.}\
  }\textbf {\bibinfo {volume} {8}},\ \bibinfo {pages} {729} (\bibinfo {year}
  {2012})}\BibitemShut {NoStop}%
\bibitem [{\citenamefont {Zhu}\ \emph {et~al.}(2011)\citenamefont {Zhu},
  \citenamefont {Yao}, \citenamefont {Zhang},\ and\ \citenamefont
  {Chang}}]{ZhuJJ2011}%
  \BibitemOpen
  \bibfield  {author} {\bibinfo {author} {\bibfnamefont {J.-J.}\ \bibnamefont
  {Zhu}}, \bibinfo {author} {\bibfnamefont {D.-X.}\ \bibnamefont {Yao}},
  \bibinfo {author} {\bibfnamefont {S.-C.}\ \bibnamefont {Zhang}}, \ and\
  \bibinfo {author} {\bibfnamefont {K.}~\bibnamefont {Chang}},\ }\href
  {\doibase 10.1103/PhysRevLett.106.097201} {\bibfield  {journal} {\bibinfo
  {journal} {Phys. Rev. Lett.}\ }\textbf {\bibinfo {volume} {106}},\ \bibinfo
  {pages} {097201} (\bibinfo {year} {2011})}\BibitemShut {NoStop}%
\bibitem [{\citenamefont {Chang}\ \emph
  {et~al.}(2013{\natexlab{c}})\citenamefont {Chang}, \citenamefont {Zhang},
  \citenamefont {Liu}, \citenamefont {Zhang}, \citenamefont {Feng},
  \citenamefont {Li}, \citenamefont {Wang}, \citenamefont {Chen}, \citenamefont
  {Dai}, \citenamefont {Fang} \emph {et~al.}}]{chang2013thin}%
  \BibitemOpen
  \bibfield  {author} {\bibinfo {author} {\bibfnamefont {C.-Z.}\ \bibnamefont
  {Chang}}, \bibinfo {author} {\bibfnamefont {J.}~\bibnamefont {Zhang}},
  \bibinfo {author} {\bibfnamefont {M.}~\bibnamefont {Liu}}, \bibinfo {author}
  {\bibfnamefont {Z.}~\bibnamefont {Zhang}}, \bibinfo {author} {\bibfnamefont
  {X.}~\bibnamefont {Feng}}, \bibinfo {author} {\bibfnamefont {K.}~\bibnamefont
  {Li}}, \bibinfo {author} {\bibfnamefont {L.-L.}\ \bibnamefont {Wang}},
  \bibinfo {author} {\bibfnamefont {X.}~\bibnamefont {Chen}}, \bibinfo {author}
  {\bibfnamefont {X.}~\bibnamefont {Dai}}, \bibinfo {author} {\bibfnamefont
  {Z.}~\bibnamefont {Fang}},  \emph {et~al.},\ }\href {\doibase
  10.1002/adma.201203493} {\bibfield  {journal} {\bibinfo  {journal} {Adv.
  Mater.}\ }\textbf {\bibinfo {volume} {25}},\ \bibinfo {pages} {1065}
  (\bibinfo {year} {2013}{\natexlab{c}})}\BibitemShut {NoStop}%
\bibitem [{\citenamefont {Chang}\ \emph {et~al.}(2014)\citenamefont {Chang},
  \citenamefont {Tang}, \citenamefont {Wang}, \citenamefont {Feng},
  \citenamefont {Li}, \citenamefont {Zhang}, \citenamefont {Wang},
  \citenamefont {Wang}, \citenamefont {Chen}, \citenamefont {Liu},
  \citenamefont {Duan}, \citenamefont {He}, \citenamefont {Ma},\ and\
  \citenamefont {Xue}}]{ChangCZ2014}%
  \BibitemOpen
  \bibfield  {author} {\bibinfo {author} {\bibfnamefont {C.-Z.}\ \bibnamefont
  {Chang}}, \bibinfo {author} {\bibfnamefont {P.}~\bibnamefont {Tang}},
  \bibinfo {author} {\bibfnamefont {Y.-L.}\ \bibnamefont {Wang}}, \bibinfo
  {author} {\bibfnamefont {X.}~\bibnamefont {Feng}}, \bibinfo {author}
  {\bibfnamefont {K.}~\bibnamefont {Li}}, \bibinfo {author} {\bibfnamefont
  {Z.}~\bibnamefont {Zhang}}, \bibinfo {author} {\bibfnamefont
  {Y.}~\bibnamefont {Wang}}, \bibinfo {author} {\bibfnamefont {L.-L.}\
  \bibnamefont {Wang}}, \bibinfo {author} {\bibfnamefont {X.}~\bibnamefont
  {Chen}}, \bibinfo {author} {\bibfnamefont {C.}~\bibnamefont {Liu}}, \bibinfo
  {author} {\bibfnamefont {W.}~\bibnamefont {Duan}}, \bibinfo {author}
  {\bibfnamefont {K.}~\bibnamefont {He}}, \bibinfo {author} {\bibfnamefont
  {X.-C.}\ \bibnamefont {Ma}}, \ and\ \bibinfo {author} {\bibfnamefont {Q.-K.}\
  \bibnamefont {Xue}},\ }\href {\doibase 10.1103/PhysRevLett.112.056801}
  {\bibfield  {journal} {\bibinfo  {journal} {Phys. Rev. Lett.}\ }\textbf
  {\bibinfo {volume} {112}},\ \bibinfo {pages} {056801} (\bibinfo {year}
  {2014})}\BibitemShut {NoStop}%
\bibitem [{\citenamefont {He}\ \emph {et~al.}(2010)\citenamefont {He},
  \citenamefont {Zhang}, \citenamefont {He}, \citenamefont {Chang},
  \citenamefont {Song}, \citenamefont {Wang}, \citenamefont {Chen},
  \citenamefont {Jia}, \citenamefont {Fang},\ and\ \citenamefont
  {Dai}}]{He2010Crossover}%
  \BibitemOpen
  \bibfield  {author} {\bibinfo {author} {\bibfnamefont {K.}~\bibnamefont
  {He}}, \bibinfo {author} {\bibfnamefont {Y.}~\bibnamefont {Zhang}}, \bibinfo
  {author} {\bibfnamefont {K.}~\bibnamefont {He}}, \bibinfo {author}
  {\bibfnamefont {C.~Z.}\ \bibnamefont {Chang}}, \bibinfo {author}
  {\bibfnamefont {C.~L.}\ \bibnamefont {Song}}, \bibinfo {author}
  {\bibfnamefont {L.~L.}\ \bibnamefont {Wang}}, \bibinfo {author}
  {\bibfnamefont {X.}~\bibnamefont {Chen}}, \bibinfo {author} {\bibfnamefont
  {J.~F.}\ \bibnamefont {Jia}}, \bibinfo {author} {\bibfnamefont
  {Z.}~\bibnamefont {Fang}}, \ and\ \bibinfo {author} {\bibfnamefont
  {X.}~\bibnamefont {Dai}},\ }\href {\doibase 10.1038/nphys1689} {\bibfield
  {journal} {\bibinfo  {journal} {Nat. Phys.}\ }\textbf {\bibinfo {volume}
  {6}},\ \bibinfo {pages} {712} (\bibinfo {year} {2010})}\BibitemShut {NoStop}%
\bibitem [{\citenamefont {Wang}\ \emph {et~al.}(2015)\citenamefont {Wang},
  \citenamefont {Lian},\ and\ \citenamefont {Zhang}}]{wang2015}%
  \BibitemOpen
  \bibfield  {author} {\bibinfo {author} {\bibfnamefont {J.}~\bibnamefont
  {Wang}}, \bibinfo {author} {\bibfnamefont {B.}~\bibnamefont {Lian}}, \ and\
  \bibinfo {author} {\bibfnamefont {S.-C.}\ \bibnamefont {Zhang}},\ }\href
  {\doibase 10.1103/PhysRevLett.115.036805} {\bibfield  {journal} {\bibinfo
  {journal} {Phys. Rev. Lett.}\ }\textbf {\bibinfo {volume} {115}},\ \bibinfo
  {pages} {036805} (\bibinfo {year} {2015})}\BibitemShut {NoStop}%
\bibitem [{\citenamefont {Zhang}\ \emph {et~al.}(2017)\citenamefont {Zhang},
  \citenamefont {Feng}, \citenamefont {Wang}, \citenamefont {Lian},
  \citenamefont {Zhang}, \citenamefont {Chang}, \citenamefont {Guo},
  \citenamefont {Ou}, \citenamefont {Feng}, \citenamefont {Zhang} \emph
  {et~al.}}]{zhang2017magnetic}%
  \BibitemOpen
  \bibfield  {author} {\bibinfo {author} {\bibfnamefont {Z.}~\bibnamefont
  {Zhang}}, \bibinfo {author} {\bibfnamefont {X.}~\bibnamefont {Feng}},
  \bibinfo {author} {\bibfnamefont {J.}~\bibnamefont {Wang}}, \bibinfo {author}
  {\bibfnamefont {B.}~\bibnamefont {Lian}}, \bibinfo {author} {\bibfnamefont
  {J.}~\bibnamefont {Zhang}}, \bibinfo {author} {\bibfnamefont
  {C.}~\bibnamefont {Chang}}, \bibinfo {author} {\bibfnamefont
  {M.}~\bibnamefont {Guo}}, \bibinfo {author} {\bibfnamefont {Y.}~\bibnamefont
  {Ou}}, \bibinfo {author} {\bibfnamefont {Y.}~\bibnamefont {Feng}}, \bibinfo
  {author} {\bibfnamefont {S.-C.}\ \bibnamefont {Zhang}},  \emph {et~al.},\
  }\href {\doibase 10.1038/nnano.2017.149} {\bibfield  {journal} {\bibinfo
  {journal} {Nat. Nano.}\ }\textbf {\bibinfo {volume} {12}},\ \bibinfo {pages}
  {953} (\bibinfo {year} {2017})}\BibitemShut {NoStop}%
\bibitem [{\citenamefont {Nomura}\ and\ \citenamefont
  {Nagaosa}(2011)}]{Namura2011}%
  \BibitemOpen
  \bibfield  {author} {\bibinfo {author} {\bibfnamefont {K.}~\bibnamefont
  {Nomura}}\ and\ \bibinfo {author} {\bibfnamefont {N.}~\bibnamefont
  {Nagaosa}},\ }\href {\doibase 10.1103/PhysRevLett.106.166802} {\bibfield
  {journal} {\bibinfo  {journal} {Phys. Rev. Lett.}\ }\textbf {\bibinfo
  {volume} {106}},\ \bibinfo {pages} {166802} (\bibinfo {year}
  {2011})}\BibitemShut {NoStop}%
\bibitem [{\citenamefont {Lu}\ \emph {et~al.}(2011)\citenamefont {Lu},
  \citenamefont {Shi},\ and\ \citenamefont {Shen}}]{LuHZ2011}%
  \BibitemOpen
  \bibfield  {author} {\bibinfo {author} {\bibfnamefont {H.-Z.}\ \bibnamefont
  {Lu}}, \bibinfo {author} {\bibfnamefont {J.}~\bibnamefont {Shi}}, \ and\
  \bibinfo {author} {\bibfnamefont {S.-Q.}\ \bibnamefont {Shen}},\ }\href
  {\doibase 10.1103/PhysRevLett.107.076801} {\bibfield  {journal} {\bibinfo
  {journal} {Phys. Rev. Lett.}\ }\textbf {\bibinfo {volume} {107}},\ \bibinfo
  {pages} {076801} (\bibinfo {year} {2011})}\BibitemShut {NoStop}%
\bibitem [{\citenamefont {Qiao}\ \emph {et~al.}(2016)\citenamefont {Qiao},
  \citenamefont {Han}, \citenamefont {Zhang}, \citenamefont {Wang},
  \citenamefont {Deng}, \citenamefont {Jiang}, \citenamefont {Yang},
  \citenamefont {Wang},\ and\ \citenamefont {Niu}}]{QiaoZH2016}%
  \BibitemOpen
  \bibfield  {author} {\bibinfo {author} {\bibfnamefont {Z.}~\bibnamefont
  {Qiao}}, \bibinfo {author} {\bibfnamefont {Y.}~\bibnamefont {Han}}, \bibinfo
  {author} {\bibfnamefont {L.}~\bibnamefont {Zhang}}, \bibinfo {author}
  {\bibfnamefont {K.}~\bibnamefont {Wang}}, \bibinfo {author} {\bibfnamefont
  {X.}~\bibnamefont {Deng}}, \bibinfo {author} {\bibfnamefont {H.}~\bibnamefont
  {Jiang}}, \bibinfo {author} {\bibfnamefont {S.~A.}\ \bibnamefont {Yang}},
  \bibinfo {author} {\bibfnamefont {J.}~\bibnamefont {Wang}}, \ and\ \bibinfo
  {author} {\bibfnamefont {Q.}~\bibnamefont {Niu}},\ }\href {\doibase
  10.1103/PhysRevLett.117.056802} {\bibfield  {journal} {\bibinfo  {journal}
  {Phys. Rev. Lett.}\ }\textbf {\bibinfo {volume} {117}},\ \bibinfo {pages}
  {056802} (\bibinfo {year} {2016})}\BibitemShut {NoStop}%
\bibitem [{\citenamefont {Chen}\ \emph {et~al.}(2019)\citenamefont {Chen},
  \citenamefont {Liu},\ and\ \citenamefont {Xie}}]{ChenCZ2019}%
  \BibitemOpen
  \bibfield  {author} {\bibinfo {author} {\bibfnamefont {C.-Z.}\ \bibnamefont
  {Chen}}, \bibinfo {author} {\bibfnamefont {H.}~\bibnamefont {Liu}}, \ and\
  \bibinfo {author} {\bibfnamefont {X.~C.}\ \bibnamefont {Xie}},\ }\href
  {\doibase 10.1103/PhysRevLett.122.026601} {\bibfield  {journal} {\bibinfo
  {journal} {Phys. Rev. Lett.}\ }\textbf {\bibinfo {volume} {122}},\ \bibinfo
  {pages} {026601} (\bibinfo {year} {2019})}\BibitemShut {NoStop}%
\bibitem [{\citenamefont {Haim}\ \emph {et~al.}(2019)\citenamefont {Haim},
  \citenamefont {Ilan},\ and\ \citenamefont {Alicea}}]{Haim2019}%
  \BibitemOpen
  \bibfield  {author} {\bibinfo {author} {\bibfnamefont {A.}~\bibnamefont
  {Haim}}, \bibinfo {author} {\bibfnamefont {R.}~\bibnamefont {Ilan}}, \ and\
  \bibinfo {author} {\bibfnamefont {J.}~\bibnamefont {Alicea}},\ }\href
  {\doibase 10.1103/PhysRevLett.123.046801} {\bibfield  {journal} {\bibinfo
  {journal} {Phys. Rev. Lett.}\ }\textbf {\bibinfo {volume} {123}},\ \bibinfo
  {pages} {046801} (\bibinfo {year} {2019})}\BibitemShut {NoStop}%
\bibitem [{\citenamefont {Xing}\ \emph {et~al.}(2018)\citenamefont {Xing},
  \citenamefont {Xu}, \citenamefont {Cheung}, \citenamefont {Sun},
  \citenamefont {Wang},\ and\ \citenamefont {Yao}}]{xing2018influence}%
  \BibitemOpen
  \bibfield  {author} {\bibinfo {author} {\bibfnamefont {Y.}~\bibnamefont
  {Xing}}, \bibinfo {author} {\bibfnamefont {F.}~\bibnamefont {Xu}}, \bibinfo
  {author} {\bibfnamefont {K.~T.}\ \bibnamefont {Cheung}}, \bibinfo {author}
  {\bibfnamefont {Q.-f.}\ \bibnamefont {Sun}}, \bibinfo {author} {\bibfnamefont
  {J.}~\bibnamefont {Wang}}, \ and\ \bibinfo {author} {\bibfnamefont
  {Y.}~\bibnamefont {Yao}},\ }\href {\doibase 10.1088/1367-2630/aab4e8}
  {\bibfield  {journal} {\bibinfo  {journal} {New J. Phys.}\ }\textbf {\bibinfo
  {volume} {20}},\ \bibinfo {pages} {043011} (\bibinfo {year}
  {2018})}\BibitemShut {NoStop}%
\bibitem [{\citenamefont {Wang}\ \emph {et~al.}(2014)\citenamefont {Wang},
  \citenamefont {Lian},\ and\ \citenamefont {Zhang}}]{WangJ2014Uni}%
  \BibitemOpen
  \bibfield  {author} {\bibinfo {author} {\bibfnamefont {J.}~\bibnamefont
  {Wang}}, \bibinfo {author} {\bibfnamefont {B.}~\bibnamefont {Lian}}, \ and\
  \bibinfo {author} {\bibfnamefont {S.-C.}\ \bibnamefont {Zhang}},\ }\href
  {\doibase 10.1103/PhysRevB.89.085106} {\bibfield  {journal} {\bibinfo
  {journal} {Phys. Rev. B}\ }\textbf {\bibinfo {volume} {89}},\ \bibinfo
  {pages} {085106} (\bibinfo {year} {2014})}\BibitemShut {NoStop}%
\bibitem [{\citenamefont {Keser}\ \emph {et~al.}(2019)\citenamefont {Keser},
  \citenamefont {Raimondi},\ and\ \citenamefont {Culcer}}]{Keser2019}%
  \BibitemOpen
  \bibfield  {author} {\bibinfo {author} {\bibfnamefont {A.~C.}\ \bibnamefont
  {Keser}}, \bibinfo {author} {\bibfnamefont {R.}~\bibnamefont {Raimondi}}, \
  and\ \bibinfo {author} {\bibfnamefont {D.}~\bibnamefont {Culcer}},\ }\href
  {\doibase 10.1103/PhysRevLett.123.126603} {\bibfield  {journal} {\bibinfo
  {journal} {Phys. Rev. Lett.}\ }\textbf {\bibinfo {volume} {123}},\ \bibinfo
  {pages} {126603} (\bibinfo {year} {2019})}\BibitemShut {NoStop}%
\bibitem [{\citenamefont {Wang}\ \emph {et~al.}(2018)\citenamefont {Wang},
  \citenamefont {Ou}, \citenamefont {Liu}, \citenamefont {Wang}, \citenamefont
  {He}, \citenamefont {Xue},\ and\ \citenamefont {Wu}}]{wang2018direct}%
  \BibitemOpen
  \bibfield  {author} {\bibinfo {author} {\bibfnamefont {W.}~\bibnamefont
  {Wang}}, \bibinfo {author} {\bibfnamefont {Y.}~\bibnamefont {Ou}}, \bibinfo
  {author} {\bibfnamefont {C.}~\bibnamefont {Liu}}, \bibinfo {author}
  {\bibfnamefont {Y.}~\bibnamefont {Wang}}, \bibinfo {author} {\bibfnamefont
  {K.}~\bibnamefont {He}}, \bibinfo {author} {\bibfnamefont {Q.-K.}\
  \bibnamefont {Xue}}, \ and\ \bibinfo {author} {\bibfnamefont
  {W.}~\bibnamefont {Wu}},\ }\href {\doibase 10.1038/s41567-018-0149-1}
  {\bibfield  {journal} {\bibinfo  {journal} {Nat. Phys.}\ }\textbf {\bibinfo
  {volume} {14}},\ \bibinfo {pages} {791} (\bibinfo {year} {2018})}\BibitemShut
  {NoStop}%
\bibitem [{\citenamefont {Lee}\ \emph {et~al.}(2015)\citenamefont {Lee},
  \citenamefont {Kim}, \citenamefont {Lee}, \citenamefont {Billinge},
  \citenamefont {Zhong}, \citenamefont {Schneeloch}, \citenamefont {Liu},
  \citenamefont {Valla}, \citenamefont {Tranquada}, \citenamefont {Gu} \emph
  {et~al.}}]{lee2015imaging}%
  \BibitemOpen
  \bibfield  {author} {\bibinfo {author} {\bibfnamefont {I.}~\bibnamefont
  {Lee}}, \bibinfo {author} {\bibfnamefont {C.~K.}\ \bibnamefont {Kim}},
  \bibinfo {author} {\bibfnamefont {J.}~\bibnamefont {Lee}}, \bibinfo {author}
  {\bibfnamefont {S.~J.}\ \bibnamefont {Billinge}}, \bibinfo {author}
  {\bibfnamefont {R.}~\bibnamefont {Zhong}}, \bibinfo {author} {\bibfnamefont
  {J.~A.}\ \bibnamefont {Schneeloch}}, \bibinfo {author} {\bibfnamefont
  {T.}~\bibnamefont {Liu}}, \bibinfo {author} {\bibfnamefont {T.}~\bibnamefont
  {Valla}}, \bibinfo {author} {\bibfnamefont {J.~M.}\ \bibnamefont
  {Tranquada}}, \bibinfo {author} {\bibfnamefont {G.}~\bibnamefont {Gu}},
  \emph {et~al.},\ }\href {\doibase 10.1073/pnas.1424322112} {\bibfield
  {journal} {\bibinfo  {journal} {PNAS}\ }\textbf {\bibinfo {volume} {112}},\
  \bibinfo {pages} {1316} (\bibinfo {year} {2015})}\BibitemShut {NoStop}%
\bibitem [{\citenamefont {Lachman}\ \emph {et~al.}(2015)\citenamefont
  {Lachman}, \citenamefont {Young}, \citenamefont {Richardella}, \citenamefont
  {Cuppens}, \citenamefont {Naren}, \citenamefont {Anahory}, \citenamefont
  {Meltzer}, \citenamefont {Kandala}, \citenamefont {Kempinger}, \citenamefont
  {Myasoedov} \emph {et~al.}}]{lachman2015vis}%
  \BibitemOpen
  \bibfield  {author} {\bibinfo {author} {\bibfnamefont {E.~O.}\ \bibnamefont
  {Lachman}}, \bibinfo {author} {\bibfnamefont {A.~F.}\ \bibnamefont {Young}},
  \bibinfo {author} {\bibfnamefont {A.}~\bibnamefont {Richardella}}, \bibinfo
  {author} {\bibfnamefont {J.}~\bibnamefont {Cuppens}}, \bibinfo {author}
  {\bibfnamefont {H.}~\bibnamefont {Naren}}, \bibinfo {author} {\bibfnamefont
  {Y.}~\bibnamefont {Anahory}}, \bibinfo {author} {\bibfnamefont {A.~Y.}\
  \bibnamefont {Meltzer}}, \bibinfo {author} {\bibfnamefont {A.}~\bibnamefont
  {Kandala}}, \bibinfo {author} {\bibfnamefont {S.}~\bibnamefont {Kempinger}},
  \bibinfo {author} {\bibfnamefont {Y.}~\bibnamefont {Myasoedov}},  \emph
  {et~al.},\ }\href {\doibase 10.1126/sciadv.1500740} {\bibfield  {journal}
  {\bibinfo  {journal} {Sci. Adv.}\ }\textbf {\bibinfo {volume} {1}},\ \bibinfo
  {pages} {e1500740} (\bibinfo {year} {2015})}\BibitemShut {NoStop}%
\bibitem [{\citenamefont {Kou}\ \emph {et~al.}(2015)\citenamefont {Kou},
  \citenamefont {Pan}, \citenamefont {Wang}, \citenamefont {Fan}, \citenamefont
  {Choi}, \citenamefont {Lee}, \citenamefont {Nie}, \citenamefont {Murata},
  \citenamefont {Shao}, \citenamefont {Zhang} \emph {et~al.}}]{kou2015metal}%
  \BibitemOpen
  \bibfield  {author} {\bibinfo {author} {\bibfnamefont {X.}~\bibnamefont
  {Kou}}, \bibinfo {author} {\bibfnamefont {L.}~\bibnamefont {Pan}}, \bibinfo
  {author} {\bibfnamefont {J.}~\bibnamefont {Wang}}, \bibinfo {author}
  {\bibfnamefont {Y.}~\bibnamefont {Fan}}, \bibinfo {author} {\bibfnamefont
  {E.~S.}\ \bibnamefont {Choi}}, \bibinfo {author} {\bibfnamefont {W.-L.}\
  \bibnamefont {Lee}}, \bibinfo {author} {\bibfnamefont {T.}~\bibnamefont
  {Nie}}, \bibinfo {author} {\bibfnamefont {K.}~\bibnamefont {Murata}},
  \bibinfo {author} {\bibfnamefont {Q.}~\bibnamefont {Shao}}, \bibinfo {author}
  {\bibfnamefont {S.-C.}\ \bibnamefont {Zhang}},  \emph {et~al.},\ }\href
  {\doibase doi.org/10.1038/ncomms9474} {\bibfield  {journal} {\bibinfo
  {journal} {Nat. Comm.}\ }\textbf {\bibinfo {volume} {6}},\ \bibinfo {pages}
  {1} (\bibinfo {year} {2015})}\BibitemShut {NoStop}%
\bibitem [{\citenamefont {Zhang}\ \emph {et~al.}(2020)\citenamefont {Zhang},
  \citenamefont {Chen}, \citenamefont {Wu}, \citenamefont {Jiang},
  \citenamefont {Liu}, \citenamefont {feng Sun},\ and\ \citenamefont
  {Xie}}]{zhang2020chiral}%
  \BibitemOpen
  \bibfield  {author} {\bibinfo {author} {\bibfnamefont {Z.-Q.}\ \bibnamefont
  {Zhang}}, \bibinfo {author} {\bibfnamefont {C.-Z.}\ \bibnamefont {Chen}},
  \bibinfo {author} {\bibfnamefont {Y.}~\bibnamefont {Wu}}, \bibinfo {author}
  {\bibfnamefont {H.}~\bibnamefont {Jiang}}, \bibinfo {author} {\bibfnamefont
  {J.}~\bibnamefont {Liu}}, \bibinfo {author} {\bibfnamefont {Q.}~\bibnamefont
  {feng Sun}}, \ and\ \bibinfo {author} {\bibfnamefont {X.~C.}\ \bibnamefont
  {Xie}},\ }\href@noop {} {\enquote {\bibinfo {title} {Chiral interface states
  and related quantized transport in disordered chern insulators},}\ }
  (\bibinfo {year} {2020}),\ \Eprint {http://arxiv.org/abs/2007.07619}
  {arXiv:2007.07619 [cond-mat.mes-hall]} \BibitemShut {NoStop}%
\bibitem [{\citenamefont {Wray}\ \emph {et~al.}(2011)\citenamefont {Wray},
  \citenamefont {Xu}, \citenamefont {Xia}, \citenamefont {Hsieh}, \citenamefont
  {Fedorov}, \citenamefont {San~Hor}, \citenamefont {Cava}, \citenamefont
  {Bansil}, \citenamefont {Lin},\ and\ \citenamefont
  {Hasan}}]{wray2011topological}%
  \BibitemOpen
  \bibfield  {author} {\bibinfo {author} {\bibfnamefont {L.~A.}\ \bibnamefont
  {Wray}}, \bibinfo {author} {\bibfnamefont {S.-Y.}\ \bibnamefont {Xu}},
  \bibinfo {author} {\bibfnamefont {Y.}~\bibnamefont {Xia}}, \bibinfo {author}
  {\bibfnamefont {D.}~\bibnamefont {Hsieh}}, \bibinfo {author} {\bibfnamefont
  {A.~V.}\ \bibnamefont {Fedorov}}, \bibinfo {author} {\bibfnamefont
  {Y.}~\bibnamefont {San~Hor}}, \bibinfo {author} {\bibfnamefont {R.~J.}\
  \bibnamefont {Cava}}, \bibinfo {author} {\bibfnamefont {A.}~\bibnamefont
  {Bansil}}, \bibinfo {author} {\bibfnamefont {H.}~\bibnamefont {Lin}}, \ and\
  \bibinfo {author} {\bibfnamefont {M.~Z.}\ \bibnamefont {Hasan}},\ }\href
  {\doibase 10.1038/nphys1838} {\bibfield  {journal} {\bibinfo  {journal} {Nat.
  Phys.}\ }\textbf {\bibinfo {volume} {7}},\ \bibinfo {pages} {32} (\bibinfo
  {year} {2011})}\BibitemShut {NoStop}%
\bibitem [{\citenamefont {Yuan}\ \emph {et~al.}(2020)\citenamefont {Yuan},
  \citenamefont {Wang}, \citenamefont {Li}, \citenamefont {Li}, \citenamefont
  {Ji}, \citenamefont {Hao}, \citenamefont {Wu}, \citenamefont {He},
  \citenamefont {Wang}, \citenamefont {Xu}, \citenamefont {Duan}, \citenamefont
  {Li},\ and\ \citenamefont {Xue}}]{YuanYH2020}%
  \BibitemOpen
  \bibfield  {author} {\bibinfo {author} {\bibfnamefont {Y.}~\bibnamefont
  {Yuan}}, \bibinfo {author} {\bibfnamefont {X.}~\bibnamefont {Wang}}, \bibinfo
  {author} {\bibfnamefont {H.}~\bibnamefont {Li}}, \bibinfo {author}
  {\bibfnamefont {J.}~\bibnamefont {Li}}, \bibinfo {author} {\bibfnamefont
  {Y.}~\bibnamefont {Ji}}, \bibinfo {author} {\bibfnamefont {Z.}~\bibnamefont
  {Hao}}, \bibinfo {author} {\bibfnamefont {Y.}~\bibnamefont {Wu}}, \bibinfo
  {author} {\bibfnamefont {K.}~\bibnamefont {He}}, \bibinfo {author}
  {\bibfnamefont {Y.}~\bibnamefont {Wang}}, \bibinfo {author} {\bibfnamefont
  {Y.}~\bibnamefont {Xu}}, \bibinfo {author} {\bibfnamefont {W.}~\bibnamefont
  {Duan}}, \bibinfo {author} {\bibfnamefont {W.}~\bibnamefont {Li}}, \ and\
  \bibinfo {author} {\bibfnamefont {Q.-K.}\ \bibnamefont {Xue}},\ }\href
  {\doibase 10.1021/acs.nanolett.0c00031} {\bibfield  {journal} {\bibinfo
  {journal} {Nano Lett.}\ }\textbf {\bibinfo {volume} {20}},\ \bibinfo {pages}
  {3271} (\bibinfo {year} {2020})}\BibitemShut {NoStop}%
\bibitem [{\citenamefont {Chen}\ \emph {et~al.}(2015)\citenamefont {Chen},
  \citenamefont {Teague}, \citenamefont {He}, \citenamefont {Kou},
  \citenamefont {Lang}, \citenamefont {Fan}, \citenamefont {Woodward},
  \citenamefont {Wang},\ and\ \citenamefont {Yeh}}]{chen2015magnetism}%
  \BibitemOpen
  \bibfield  {author} {\bibinfo {author} {\bibfnamefont {C.}~\bibnamefont
  {Chen}}, \bibinfo {author} {\bibfnamefont {M.}~\bibnamefont {Teague}},
  \bibinfo {author} {\bibfnamefont {L.}~\bibnamefont {He}}, \bibinfo {author}
  {\bibfnamefont {X.}~\bibnamefont {Kou}}, \bibinfo {author} {\bibfnamefont
  {M.}~\bibnamefont {Lang}}, \bibinfo {author} {\bibfnamefont {W.}~\bibnamefont
  {Fan}}, \bibinfo {author} {\bibfnamefont {N.}~\bibnamefont {Woodward}},
  \bibinfo {author} {\bibfnamefont {K.}~\bibnamefont {Wang}}, \ and\ \bibinfo
  {author} {\bibfnamefont {N.}~\bibnamefont {Yeh}},\ }\href {\doibase
  10.1088/1367-2630/17/11/113042} {\bibfield  {journal} {\bibinfo  {journal}
  {New J. Phys.}\ }\textbf {\bibinfo {volume} {17}},\ \bibinfo {pages} {113042}
  (\bibinfo {year} {2015})}\BibitemShut {NoStop}%
\bibitem [{\citenamefont {Liao}\ \emph {et~al.}(2015)\citenamefont {Liao},
  \citenamefont {Ou}, \citenamefont {Feng}, \citenamefont {Yang}, \citenamefont
  {Lin}, \citenamefont {Yang}, \citenamefont {Wu}, \citenamefont {He},
  \citenamefont {Ma}, \citenamefont {Xue},\ and\ \citenamefont
  {Li}}]{LiaoJ2015}%
  \BibitemOpen
  \bibfield  {author} {\bibinfo {author} {\bibfnamefont {J.}~\bibnamefont
  {Liao}}, \bibinfo {author} {\bibfnamefont {Y.}~\bibnamefont {Ou}}, \bibinfo
  {author} {\bibfnamefont {X.}~\bibnamefont {Feng}}, \bibinfo {author}
  {\bibfnamefont {S.}~\bibnamefont {Yang}}, \bibinfo {author} {\bibfnamefont
  {C.}~\bibnamefont {Lin}}, \bibinfo {author} {\bibfnamefont {W.}~\bibnamefont
  {Yang}}, \bibinfo {author} {\bibfnamefont {K.}~\bibnamefont {Wu}}, \bibinfo
  {author} {\bibfnamefont {K.}~\bibnamefont {He}}, \bibinfo {author}
  {\bibfnamefont {X.}~\bibnamefont {Ma}}, \bibinfo {author} {\bibfnamefont
  {Q.-K.}\ \bibnamefont {Xue}}, \ and\ \bibinfo {author} {\bibfnamefont
  {Y.}~\bibnamefont {Li}},\ }\href {\doibase 10.1103/PhysRevLett.114.216601}
  {\bibfield  {journal} {\bibinfo  {journal} {Phys. Rev. Lett.}\ }\textbf
  {\bibinfo {volume} {114}},\ \bibinfo {pages} {216601} (\bibinfo {year}
  {2015})}\BibitemShut {NoStop}%
\bibitem [{\citenamefont {Li}\ \emph {et~al.}(2009)\citenamefont {Li},
  \citenamefont {Chu}, \citenamefont {Jain},\ and\ \citenamefont
  {Shen}}]{Li2009}%
  \BibitemOpen
  \bibfield  {author} {\bibinfo {author} {\bibfnamefont {J.}~\bibnamefont
  {Li}}, \bibinfo {author} {\bibfnamefont {R.-L.}\ \bibnamefont {Chu}},
  \bibinfo {author} {\bibfnamefont {J.~K.}\ \bibnamefont {Jain}}, \ and\
  \bibinfo {author} {\bibfnamefont {S.-Q.}\ \bibnamefont {Shen}},\ }\href
  {\doibase 10.1103/PhysRevLett.102.136806} {\bibfield  {journal} {\bibinfo
  {journal} {Phys. Rev. Lett.}\ }\textbf {\bibinfo {volume} {102}},\ \bibinfo
  {pages} {136806} (\bibinfo {year} {2009})}\BibitemShut {NoStop}%
\bibitem [{\citenamefont {Groth}\ \emph {et~al.}(2009)\citenamefont {Groth},
  \citenamefont {Wimmer}, \citenamefont {Akhmerov}, \citenamefont
  {Tworzyd\l{}o},\ and\ \citenamefont {Beenakker}}]{Groth2009}%
  \BibitemOpen
  \bibfield  {author} {\bibinfo {author} {\bibfnamefont {C.~W.}\ \bibnamefont
  {Groth}}, \bibinfo {author} {\bibfnamefont {M.}~\bibnamefont {Wimmer}},
  \bibinfo {author} {\bibfnamefont {A.~R.}\ \bibnamefont {Akhmerov}}, \bibinfo
  {author} {\bibfnamefont {J.}~\bibnamefont {Tworzyd\l{}o}}, \ and\ \bibinfo
  {author} {\bibfnamefont {C.~W.~J.}\ \bibnamefont {Beenakker}},\ }\href
  {\doibase 10.1103/PhysRevLett.103.196805} {\bibfield  {journal} {\bibinfo
  {journal} {Phys. Rev. Lett.}\ }\textbf {\bibinfo {volume} {103}},\ \bibinfo
  {pages} {196805} (\bibinfo {year} {2009})}\BibitemShut {NoStop}%
\bibitem [{\citenamefont {Jiang}\ \emph {et~al.}(2009)\citenamefont {Jiang},
  \citenamefont {Wang}, \citenamefont {Sun},\ and\ \citenamefont
  {Xie}}]{Jiang2009}%
  \BibitemOpen
  \bibfield  {author} {\bibinfo {author} {\bibfnamefont {H.}~\bibnamefont
  {Jiang}}, \bibinfo {author} {\bibfnamefont {L.}~\bibnamefont {Wang}},
  \bibinfo {author} {\bibfnamefont {Q.-f.}\ \bibnamefont {Sun}}, \ and\
  \bibinfo {author} {\bibfnamefont {X.~C.}\ \bibnamefont {Xie}},\ }\href
  {\doibase 10.1103/PhysRevB.80.165316} {\bibfield  {journal} {\bibinfo
  {journal} {Phys. Rev. B}\ }\textbf {\bibinfo {volume} {80}},\ \bibinfo
  {pages} {165316} (\bibinfo {year} {2009})}\BibitemShut {NoStop}%
\bibitem [{\citenamefont {Yamakage}\ \emph {et~al.}(2013)\citenamefont
  {Yamakage}, \citenamefont {Nomura}, \citenamefont {Imura},\ and\
  \citenamefont {Kuramoto}}]{Yamakage2013}%
  \BibitemOpen
  \bibfield  {author} {\bibinfo {author} {\bibfnamefont {A.}~\bibnamefont
  {Yamakage}}, \bibinfo {author} {\bibfnamefont {K.}~\bibnamefont {Nomura}},
  \bibinfo {author} {\bibfnamefont {K.-I.}\ \bibnamefont {Imura}}, \ and\
  \bibinfo {author} {\bibfnamefont {Y.}~\bibnamefont {Kuramoto}},\ }\href
  {\doibase 10.1103/PhysRevB.87.205141} {\bibfield  {journal} {\bibinfo
  {journal} {Phys. Rev. B}\ }\textbf {\bibinfo {volume} {87}},\ \bibinfo
  {pages} {205141} (\bibinfo {year} {2013})}\BibitemShut {NoStop}%
\bibitem [{\citenamefont {Song}\ \emph {et~al.}(2012)\citenamefont {Song},
  \citenamefont {Liu}, \citenamefont {Jiang}, \citenamefont {Sun},\ and\
  \citenamefont {Xie}}]{SongJ2012}%
  \BibitemOpen
  \bibfield  {author} {\bibinfo {author} {\bibfnamefont {J.}~\bibnamefont
  {Song}}, \bibinfo {author} {\bibfnamefont {H.}~\bibnamefont {Liu}}, \bibinfo
  {author} {\bibfnamefont {H.}~\bibnamefont {Jiang}}, \bibinfo {author}
  {\bibfnamefont {Q.-f.}\ \bibnamefont {Sun}}, \ and\ \bibinfo {author}
  {\bibfnamefont {X.~C.}\ \bibnamefont {Xie}},\ }\href {\doibase
  10.1103/PhysRevB.85.195125} {\bibfield  {journal} {\bibinfo  {journal} {Phys.
  Rev. B}\ }\textbf {\bibinfo {volume} {85}},\ \bibinfo {pages} {195125}
  (\bibinfo {year} {2012})}\BibitemShut {NoStop}%
\bibitem [{\citenamefont {Sheng}\ \emph {et~al.}(2006)\citenamefont {Sheng},
  \citenamefont {Weng}, \citenamefont {Sheng},\ and\ \citenamefont
  {Haldane}}]{ShengDN2006}%
  \BibitemOpen
  \bibfield  {author} {\bibinfo {author} {\bibfnamefont {D.~N.}\ \bibnamefont
  {Sheng}}, \bibinfo {author} {\bibfnamefont {Z.~Y.}\ \bibnamefont {Weng}},
  \bibinfo {author} {\bibfnamefont {L.}~\bibnamefont {Sheng}}, \ and\ \bibinfo
  {author} {\bibfnamefont {F.~D.~M.}\ \bibnamefont {Haldane}},\ }\href
  {\doibase 10.1103/PhysRevLett.97.036808} {\bibfield  {journal} {\bibinfo
  {journal} {Phys. Rev. Lett.}\ }\textbf {\bibinfo {volume} {97}},\ \bibinfo
  {pages} {036808} (\bibinfo {year} {2006})}\BibitemShut {NoStop}%
\bibitem [{Not()}]{Note_parameter}%
  \BibitemOpen
  \href@noop {} {\ }\bibinfo {note} {The parameters in simulations are
  $v_F=3.07/2.36$ eV \AA, $m_0=44/-29$ meV, $B=37.3/12.9$ eV \AA$^2$ for (Bi,
  Sb)$_2$Te$_3$ thin film with thickness of 3QLs/4QLs~\cite{wang2015}. Solid
  and dash lines correspond to phase boundaries calculated from the
  self-consistent Born approximation for the upper and lower block of the
  Hamiltonian. The blue and purple lines show
  $|\overline{\mu}^{u/l}|=\overline{m}_{0}^{u/l}$ and
  $|\overline{\mu}^{u/l}|=-\overline{m}_{0}^{u/l}$, respectively.}\BibitemShut
  {Stop}%
\bibitem [{Sup()}]{SuppMater}%
  \BibitemOpen
  \href@noop {} {\ }\bibinfo {note} {See Supplemental Material for details
  about the Born approximation, lead effect, and the topological phase
  transition driven by disorder for 4QLs.}\BibitemShut {Stop}%
\bibitem [{\citenamefont {Landauer}(1970)}]{Landauer1970}%
  \BibitemOpen
  \bibfield  {author} {\bibinfo {author} {\bibfnamefont {R.}~\bibnamefont
  {Landauer}},\ }\href {\doibase 10.1080/14786437008238472} {\bibfield
  {journal} {\bibinfo  {journal} {Philos. Mag.}\ }\textbf {\bibinfo {volume}
  {21}},\ \bibinfo {pages} {863} (\bibinfo {year} {1970})}\BibitemShut
  {NoStop}%
\bibitem [{\citenamefont {B\"uttiker}(1988)}]{Buttiker1988}%
  \BibitemOpen
  \bibfield  {author} {\bibinfo {author} {\bibfnamefont {M.}~\bibnamefont
  {B\"uttiker}},\ }\href {\doibase 10.1103/PhysRevB.38.9375} {\bibfield
  {journal} {\bibinfo  {journal} {Phys. Rev. B}\ }\textbf {\bibinfo {volume}
  {38}},\ \bibinfo {pages} {9375} (\bibinfo {year} {1988})}\BibitemShut
  {NoStop}%
\bibitem [{\citenamefont {Liu}\ \emph {et~al.}(2010)\citenamefont {Liu},
  \citenamefont {Zhang}, \citenamefont {Yan}, \citenamefont {Qi}, \citenamefont
  {Frauenheim}, \citenamefont {Dai}, \citenamefont {Fang},\ and\ \citenamefont
  {Zhang}}]{LiuCX2010-Osci}%
  \BibitemOpen
  \bibfield  {author} {\bibinfo {author} {\bibfnamefont {C.-X.}\ \bibnamefont
  {Liu}}, \bibinfo {author} {\bibfnamefont {H.}~\bibnamefont {Zhang}}, \bibinfo
  {author} {\bibfnamefont {B.}~\bibnamefont {Yan}}, \bibinfo {author}
  {\bibfnamefont {X.-L.}\ \bibnamefont {Qi}}, \bibinfo {author} {\bibfnamefont
  {T.}~\bibnamefont {Frauenheim}}, \bibinfo {author} {\bibfnamefont
  {X.}~\bibnamefont {Dai}}, \bibinfo {author} {\bibfnamefont {Z.}~\bibnamefont
  {Fang}}, \ and\ \bibinfo {author} {\bibfnamefont {S.-C.}\ \bibnamefont
  {Zhang}},\ }\href {\doibase 10.1103/PhysRevB.81.041307} {\bibfield  {journal}
  {\bibinfo  {journal} {Phys. Rev. B}\ }\textbf {\bibinfo {volume} {81}},\
  \bibinfo {pages} {041307} (\bibinfo {year} {2010})}\BibitemShut {NoStop}%
\bibitem [{not()}]{noteSpinChern}%
  \BibitemOpen
  \href@noop {} {\ }\bibinfo {note} {The spin-Chern number is defined in
  orbital space instead of real spin space.}\BibitemShut {Stop}%
\bibitem [{Note1()}]{Note1}%
  \BibitemOpen
  \bibinfo {note} {To discuss this strong disorder phase boundary, a scaling
  analysis as done in Ref.~\cite {Groth2009} can be performed. This is,
  however, beyond the scope of this work.}\BibitemShut {Stop}%
\bibitem [{\citenamefont {Lee}\ and\ \citenamefont
  {Ramakrishnan}(1985)}]{PatrickLee1985}%
  \BibitemOpen
  \bibfield  {author} {\bibinfo {author} {\bibfnamefont {P.~A.}\ \bibnamefont
  {Lee}}\ and\ \bibinfo {author} {\bibfnamefont {T.~V.}\ \bibnamefont
  {Ramakrishnan}},\ }\href {\doibase 10.1103/RevModPhys.57.287} {\bibfield
  {journal} {\bibinfo  {journal} {Rev. Mod. Phys.}\ }\textbf {\bibinfo {volume}
  {57}},\ \bibinfo {pages} {287} (\bibinfo {year} {1985})}\BibitemShut
  {NoStop}%
\bibitem [{\citenamefont {Su}\ \emph {et~al.}(2016)\citenamefont {Su},
  \citenamefont {Avishai},\ and\ \citenamefont {Wang}}]{Suying2016}%
  \BibitemOpen
  \bibfield  {author} {\bibinfo {author} {\bibfnamefont {Y.}~\bibnamefont
  {Su}}, \bibinfo {author} {\bibfnamefont {Y.}~\bibnamefont {Avishai}}, \ and\
  \bibinfo {author} {\bibfnamefont {X.~R.}\ \bibnamefont {Wang}},\ }\href
  {\doibase 10.1103/PhysRevB.93.214206} {\bibfield  {journal} {\bibinfo
  {journal} {Phys. Rev. B}\ }\textbf {\bibinfo {volume} {93}},\ \bibinfo
  {pages} {214206} (\bibinfo {year} {2016})}\BibitemShut {NoStop}%
\bibitem [{\citenamefont {Chen}\ \emph {et~al.}(2017)\citenamefont {Chen},
  \citenamefont {Xu},\ and\ \citenamefont {Zhou}}]{ChenRui2017}%
  \BibitemOpen
  \bibfield  {author} {\bibinfo {author} {\bibfnamefont {R.}~\bibnamefont
  {Chen}}, \bibinfo {author} {\bibfnamefont {D.-H.}\ \bibnamefont {Xu}}, \ and\
  \bibinfo {author} {\bibfnamefont {B.}~\bibnamefont {Zhou}},\ }\href {\doibase
  10.1103/PhysRevB.96.205304} {\bibfield  {journal} {\bibinfo  {journal} {Phys.
  Rev. B}\ }\textbf {\bibinfo {volume} {96}},\ \bibinfo {pages} {205304}
  (\bibinfo {year} {2017})}\BibitemShut {NoStop}%
\bibitem [{wor()}]{work_progress}%
  \BibitemOpen
  \href@noop {} {\ }\bibinfo {note} {Takuya Okugawa, Peizhe Tang, Angel Rubio,
  and Dante M. Kennes, work in progress.}\BibitemShut {Stop}%
\end{thebibliography}%


\providecommand{\noopsort}[1]{}\providecommand{\singleletter}[1]{#1}%
\begin{thebibliography}{5}%
\makeatletter
\providecommand \@ifxundefined [1]{%
 \@ifx{#1\undefined}
}%
\providecommand \@ifnum [1]{%
 \ifnum #1\expandafter \@firstoftwo
 \else \expandafter \@secondoftwo
 \fi
}%
\providecommand \@ifx [1]{%
 \ifx #1\expandafter \@firstoftwo
 \else \expandafter \@secondoftwo
 \fi
}%
\providecommand \natexlab [1]{#1}%
\providecommand \enquote  [1]{``#1''}%
\providecommand \bibnamefont  [1]{#1}%
\providecommand \bibfnamefont [1]{#1}%
\providecommand \citenamefont [1]{#1}%
\providecommand \href@noop [0]{\@secondoftwo}%
\providecommand \href [0]{\begingroup \@sanitize@url \@href}%
\providecommand \@href[1]{\@@startlink{#1}\@@href}%
\providecommand \@@href[1]{\endgroup#1\@@endlink}%
\providecommand \@sanitize@url [0]{\catcode `\\12\catcode `\$12\catcode
  `\&12\catcode `\#12\catcode `\^12\catcode `\_12\catcode `\%12\relax}%
\providecommand \@@startlink[1]{}%
\providecommand \@@endlink[0]{}%
\providecommand \url  [0]{\begingroup\@sanitize@url \@url }%
\providecommand \@url [1]{\endgroup\@href {#1}{\urlprefix }}%
\providecommand \urlprefix  [0]{URL }%
\providecommand \Eprint [0]{\href }%
\providecommand \doibase [0]{http://dx.doi.org/}%
\providecommand \selectlanguage [0]{\@gobble}%
\providecommand \bibinfo  [0]{\@secondoftwo}%
\providecommand \bibfield  [0]{\@secondoftwo}%
\providecommand \translation [1]{[#1]}%
\providecommand \BibitemOpen [0]{}%
\providecommand \bibitemStop [0]{}%
\providecommand \bibitemNoStop [0]{.\EOS\space}%
\providecommand \EOS [0]{\spacefactor3000\relax}%
\providecommand \BibitemShut  [1]{\csname bibitem#1\endcsname}%
\let\auto@bib@innerbib\@empty
%</preamble>
\bibitem [{\citenamefont {Groth}\ \emph {et~al.}(2009)\citenamefont {Groth},
  \citenamefont {Wimmer}, \citenamefont {Akhmerov}, \citenamefont
  {Tworzyd\l{}o},\ and\ \citenamefont {Beenakker}}]{Groth2009}%
  \BibitemOpen
  \bibfield  {author} {\bibinfo {author} {\bibfnamefont {C.~W.}\ \bibnamefont
  {Groth}}, \bibinfo {author} {\bibfnamefont {M.}~\bibnamefont {Wimmer}},
  \bibinfo {author} {\bibfnamefont {A.~R.}\ \bibnamefont {Akhmerov}}, \bibinfo
  {author} {\bibfnamefont {J.}~\bibnamefont {Tworzyd\l{}o}}, \ and\ \bibinfo
  {author} {\bibfnamefont {C.~W.~J.}\ \bibnamefont {Beenakker}},\ }\href
  {\doibase 10.1103/PhysRevLett.103.196805} {\bibfield  {journal} {\bibinfo
  {journal} {Phys. Rev. Lett.}\ }\textbf {\bibinfo {volume} {103}},\ \bibinfo
  {pages} {196805} (\bibinfo {year} {2009})}\BibitemShut {NoStop}%
\bibitem [{Not()}]{Note_parameter}%
  \BibitemOpen
  \href@noop {} {\ }\bibinfo {note} {The parameters in simulations are
  $v_F=3.07/2.36$ eV \AA, $m_0=44/-29$ meV, $B=37.3/12.9$ eV \AA$^2$ for (Bi,
  Sb)$_2$Te$_3$ thin film with thickness of 3QLs/4QLs~\cite{wang2015}. Solid
  and dash lines correspond to phase boundaries calculated from the
  self-consistent Born approximation for the upper and lower block of the
  Hamiltonian. The blue and purple lines show
  $|\overline{\mu}^{u/l}|=\overline{m}_{0}^{u/l}$ and
  $|\overline{\mu}^{u/l}|=-\overline{m}_{0}^{u/l}$, respectively.}\BibitemShut
  {Stop}%
\bibitem [{\citenamefont {Li}\ \emph {et~al.}(2009)\citenamefont {Li},
  \citenamefont {Chu}, \citenamefont {Jain},\ and\ \citenamefont
  {Shen}}]{Li2009}%
  \BibitemOpen
  \bibfield  {author} {\bibinfo {author} {\bibfnamefont {J.}~\bibnamefont
  {Li}}, \bibinfo {author} {\bibfnamefont {R.-L.}\ \bibnamefont {Chu}},
  \bibinfo {author} {\bibfnamefont {J.~K.}\ \bibnamefont {Jain}}, \ and\
  \bibinfo {author} {\bibfnamefont {S.-Q.}\ \bibnamefont {Shen}},\ }\href
  {\doibase 10.1103/PhysRevLett.102.136806} {\bibfield  {journal} {\bibinfo
  {journal} {Phys. Rev. Lett.}\ }\textbf {\bibinfo {volume} {102}},\ \bibinfo
  {pages} {136806} (\bibinfo {year} {2009})}\BibitemShut {NoStop}%
\bibitem [{\citenamefont {Jiang}\ \emph {et~al.}(2009)\citenamefont {Jiang},
  \citenamefont {Wang}, \citenamefont {Sun},\ and\ \citenamefont
  {Xie}}]{Jiang2009}%
  \BibitemOpen
  \bibfield  {author} {\bibinfo {author} {\bibfnamefont {H.}~\bibnamefont
  {Jiang}}, \bibinfo {author} {\bibfnamefont {L.}~\bibnamefont {Wang}},
  \bibinfo {author} {\bibfnamefont {Q.-f.}\ \bibnamefont {Sun}}, \ and\
  \bibinfo {author} {\bibfnamefont {X.~C.}\ \bibnamefont {Xie}},\ }\href
  {\doibase 10.1103/PhysRevB.80.165316} {\bibfield  {journal} {\bibinfo
  {journal} {Phys. Rev. B}\ }\textbf {\bibinfo {volume} {80}},\ \bibinfo
  {pages} {165316} (\bibinfo {year} {2009})}\BibitemShut {NoStop}%
\bibitem [{\citenamefont {Wang}\ \emph {et~al.}(2015)\citenamefont {Wang},
  \citenamefont {Lian},\ and\ \citenamefont {Zhang}}]{wang2015}%
  \BibitemOpen
  \bibfield  {author} {\bibinfo {author} {\bibfnamefont {J.}~\bibnamefont
  {Wang}}, \bibinfo {author} {\bibfnamefont {B.}~\bibnamefont {Lian}}, \ and\
  \bibinfo {author} {\bibfnamefont {S.-C.}\ \bibnamefont {Zhang}},\ }\href
  {\doibase 10.1103/PhysRevLett.115.036805} {\bibfield  {journal} {\bibinfo
  {journal} {Phys. Rev. Lett.}\ }\textbf {\bibinfo {volume} {115}},\ \bibinfo
  {pages} {036805} (\bibinfo {year} {2015})}\BibitemShut {NoStop}%
\end{thebibliography}%

\end{document}

\title{Supplemental Materials: Topological Phase Transitions Induced by Disorder in Magnetically Doped Topological Insulator (Bi, Sb)$_2$Te$_3$ Thin Films}

\author{Takuya Okugawa}
%\email{okugawa@physik.rwth-aachen.de}
\affiliation{Institut f\"ur Theorie der Statistischen Physik, RWTH Aachen, 
52056 Aachen, Germany and JARA - Fundamentals of Future Information Technology.}

\author{Peizhe Tang}
\email{peizhet@buaa.edu.cn}
\affiliation{School of Materials Science and Engineering, Beihang University, Beijing 100191, P. R. China.}
\affiliation{Max Planck Institute for the Structure and Dynamics of Matter, Center for Free Electron Laser Science, 22761 Hamburg, Germany.}

\author{Angel Rubio}
\affiliation{Max Planck Institute for the Structure and Dynamics of Matter, Center for Free Electron Laser Science, 22761 Hamburg, Germany.}
\affiliation{Center for Computational Quantum Physics, Simons Foundation Flatiron Institute, New York, NY 10010 USA}
\affiliation{Nano-Bio Spectroscopy Group,  Departamento de Fisica de Materiales, Universidad del Pa\'is Vasco, UPV/EHU- 20018 San Sebasti\'an, Spain}

\author{Dante M.\ Kennes}
\email{Dante.Kennes@rwth-aachen.de}
\affiliation{Institut f\"ur Theorie der Statistischen Physik, RWTH Aachen, 
52056 Aachen, Germany and JARA - Fundamentals of Future Information Technology.}
\affiliation{Max Planck Institute for the Structure and Dynamics of Matter, Center for Free Electron Laser Science, 22761 Hamburg, Germany.}

%\begin{abstract} 
%brabra we show a variaty of topologivcal phase transition brarbra
%\end{abstract}

\pacs{} 
\date{\today} 
\maketitle

\section{Real Space Formulation of the Model}
The lattice version of the  Hamiltonian of Eq.~(1) of the main text is given by  
\begin{equation}
H_0=
\begin{pmatrix}
h(\bm{k}) +gM\sigma_z & 0 \\
0 & h^*(\bm{k})-gM\sigma_z \label{hamiltonian_s} \\
\end{pmatrix}
\end{equation}
\begin{align}
h(\bm{k})&=\bm{d(\bm{k})} \cdot \bm{\sigma} \\
\bm{d(\bm{k})}&=[v_F a^{-1}\sin(k_y a), -v_F a^{-1}\sin(k_x a), m(\bm{k})] \\
m(\bm{k})&=m_0+2B a^{-2}[2- \cos(k_x a)+\cos(k_y a)],
\end{align}
where $a$ is the lattice constant and we take $a=2$ nm for all of our calculations. Obviously, Eq.~\eqref{hamiltonian_s} reduces to Eq.~(1) of the main text if expanded to second order around $k_x=k_y=0$,  i.e. around the $\Gamma$-point. In the momentum space, We define the respective dispersions $E^{u/l}=\pm \sqrt{(v_F k)^2 + m^{u/l} (\bm{k})^2}$, where $m^{u/l} (\bm{k})=m(\bm{k}) \pm gM$.

Furthermore, in order to model disorder effects, we transform  Eq.~\eqref{hamiltonian_s} to real space obtaining a square lattice with:
\begin{align}
H_0&=\Sigma_{ii'jj'}H_{ii'jj'}c_{i'j'}^{\dagger}c_{ij} \label{lattice_real}\\  
H_{ii'jj'}&= M\delta_{ii'}\delta_{jj'} + P_x \delta_{i+1 i'}\delta_{jj'} +P_x^{\dagger}  \delta_{i i'+1}\delta_{jj'} \notag \\
&+ P_y \delta_{i i'}\delta_{j+1 j'} +P_y^{\dagger} \delta_{i i'}\delta_{j j'+1} \\
M&=(m_0+4B) \Gamma_1 + gM \Gamma_2 \\
P_x&= -B \Gamma_1 - \frac{v_F}{2\mathrm{i}} \Gamma_4 \\
P_y&= -B \Gamma_1 + \frac{v_F}{2\mathrm{i}} \Gamma_3 
\end{align}
where $i$ and $j$ are the sites' x and y coordinates, respectively. $\Gamma^{1,2,3,4}$ are defined as follows:
\begin{equation}
\Gamma_1=
\begin{pmatrix}
\sigma_z & 0 \\
0 & \sigma_z  
\end{pmatrix}
\end{equation}
\begin{equation}
\Gamma_2=
\begin{pmatrix}
\sigma_z & 0 \\
0 & -\sigma_z  
\end{pmatrix}
\end{equation}
\begin{equation}
\Gamma_3=
\begin{pmatrix}
\sigma_x & 0 \\
0 & \sigma_x  
\end{pmatrix}
\end{equation}
\begin{equation}
\Gamma_4=
\begin{pmatrix}
\sigma_y & 0 \\
0 & -\sigma_y  
\end{pmatrix}
\end{equation}
%$\Gamma_1=\sigma_z \otimes I$, $\Gamma_2=\sigma_z \otimes \tau_z$, $\Gamma_3=\sigma_x \otimes I$, $\Gamma_4=\sigma_y \otimes \tau_z$, %where $\sigma^{i}$ act on spin space ($\uparrow$, $\downarrow$) and $\tau^{i}$ acts on (anti-)symmetric space ($+$, $-$). 
The impurities are modeled as randomly distributed on-site potentials drawn from a uniform distribution between $[-W/2, W/2]$. The term to be added to $H_{ii'jj'}$ is explicitly given as:
\begin{align}
W_{ii'jj'}&=\delta_{ii'}\delta_{jj'}W_{ij}\\
W_{ij}&=
\begin{pmatrix}
W_{ij,+ \uparrow} & 0 &0 & 0 \\
0 & W_{ij,- \downarrow} &0 & 0 \\
0 & 0 & W_{ij,+ \downarrow} & 0 \\
0 & 0 &0 & W_{ij,- \uparrow}. 
\end{pmatrix}
\end{align}
In our work, since non-magnetic impurity are considered, $W_{ij,+ \uparrow}$ = $W_{ij,+ \downarrow}$ and $W_{ij,- \uparrow}$ = $W_{ij,- \downarrow}$ are used. Adding the disorder amounts to substituting $M \rightarrow M + W_{ij}$ in Eq.~\eqref{lattice_real} and we employ this Hamiltonian for the central region described in the main text with the number of lattice sites being $N_x=L_x/a$ and $N_y=L_y/a$. For the left and right leads, the clean Hamiltonian of Eq.~\eqref{lattice_real} without any disorder is used.% for the left/right lead.% as a clean material.   

\section{Self-consistent Born approximation}

First, we calculate the static self-energy $\Sigma^{u/l}$ defined by $(E_F - H_{0}^{u/l} - \Sigma^{u/l})^{-1} = \left\langle (E_F - H^{u/l})^{-1} \right\rangle$, where $\left\langle \cdots \right\rangle$ and $H_{0}^{u/l} (H^{u/l})$ denote the disorder average and the upper/lower Block of %Eq.~(1) in the main text, respectively. 
\eqref{hamiltonian_s} without(with) disorder potential, respectively. Like the Hamiltonian, the self-energy can be decomposed into blocks $\Sigma^{u/l}=\Sigma_{0}^{u/l}\sigma_{0}+\Sigma_{x}^{u/l}\sigma_{x}+\Sigma_{y}^{u/l}\sigma_{y}+\Sigma_{z}^{u/l}\sigma_{z}$, from which the renormalized mass and chemical potential can be defined as   
%\begin{align}
$\overline{m}_{0}^{u/l}=m_{0}^{u/l}+ \Sigma_{z}^{u/l}$% \label{mbar} \\
 and 
$\overline{\mu}^{u/l}=E_{F}- \Sigma_{0}^{u/l},$% \label{mubar}     
%\end{align}
where $m_{0}^{u/l}=m_{0} \pm gM$ for the upper/lower block. For systems with a finite exchange field the time reversal symmetry is broken and the renormalization is different for the upper/lower block $\Sigma^{u}\neq\Sigma^{l}$. The self-consistent Born approximation, yields
\begin{align}
\Sigma^{u/l}&=\frac{W^2a^2}{48\pi^2} \lim\limits_{\eta\to0^+}\int_{\rm FBZ} d \bm{k} [E_{F} + i \eta - H_{0}^{u/l} - \Sigma^{u/l}]^{-1},    \label{born}
\end{align}
with FBZ denoting the first Brillouin zone. Within the Born approximation, a sign change of the effective mass term $\overline{m}_{0}^{u/l}$ signals a topological phase transition (TPT) induced by the disorder, however, to observe the topologically protected quantized conductance, the effective chemical potential renormalized by disorder must additionally be located in the bulk gap, such that the current is carried exclusively by the edge states. Consequently the phase boundary, reflecting the conductive behavior of the system, is defined by the additional condition $|\overline{\mu}^{u/l}|=-\overline{m}_{0}^{u/l}$ for $\overline{m}_{0}^{u/l}<0$ and  $|\overline{\mu}^{u/l}|=\overline{m}_{0}^{u/l}$ for $\overline{m}_{0}^{u/l}>0$.

\section{Non-self-consistent solution of the Born Self-energy}
Neglecting the feedback of $ \Sigma^{u/l}$ on the right hand side of  Eq.~\eqref{born} and keeping only the logarithmically divergent part of the integral~\cite{Groth2009}, we can obtain a closed form expression  of the renormalized mass term $\overline{m}_{0}^{u/l}$ and chemical potential $\overline{\mu}^{u/l}$:
\begin{align}
\overline{m}_{0}^{u/l}&=m_{0}^{u/l} - \frac{W^2 a^2}{48 \pi}  \frac{1}{B} \ln \left| \frac{B^2}{E_{F}^2-(m_{0}^{u/l})^2} \left(   \frac{\pi}{a}  \right)^4    \right| \label{close}  \\
\bar{\mu}&=E_{F}.   
\end{align}
Eq.~\eqref{close} shows that the disorder effects only the mass and not the chemical potential within the non-self-consistent approach.  The change in mass $\delta m_{0}^{u/l}=\overline{m}_{0}^{u/l}-m_{0}^{u/l}$  is found to always be negative for the parameters used in our calculations. This explains the tendency of disorder to promote topologically non-trivial bands in our study of magnetically doped topological insulator (Bi, Sb)$_2$Te$_3$ thin films.

\begin{figure}[t]
{\includegraphics[width=0.9\columnwidth,clip]{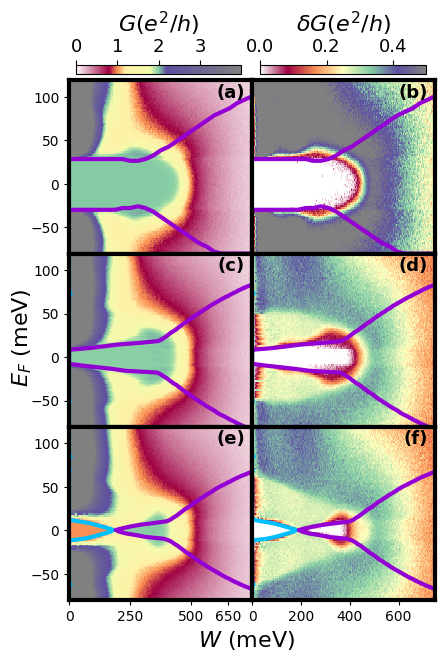}}
\caption{Left (a), (c), (e) and right (b), (d), (f) panels show the average conductance $G$ and the corresponding standard deviation $\delta G$ as a function of disorder strength $W$ and Fermi energy $E_F$ for 4QLs (Bi, Sb)$_2$Te$_3$ thin film with and without magnetic doping. The exchange field $gM$ is taken to be (a), (b): 0 meV, (c), (d): 20 meV, and (e), (f): 40 meV. The color lines stand for phase boundaries from self-consistent Born approximation. %~\cite{Note_parameter}. %Other parameters are $v_F=2360$ meV \AA, $m_0=-29$ meV, $B=12900$ meV \AA$^2$ for all, which are corresponding to 4QLs~\cite{wang2015}.
%Negative $m$ indicates nontrivial insulator regime. 
Details of the colored lines and calculation parameters are shown in Ref.~\cite{Note_parameter}. In all calculations, the system sizes are set to $L_x=400a$ and $L_y=100a$ and averages over 500 random configuration are performed.}
\label{4QL}
\end{figure}

\section{Phase Diagram of 4QLs as a function of disorder strength and Fermi energy}
In Fig.~\ref{4QL} we analyze the 4QLs case. Without exchange field $gM=0$, shown in panels (a) and (b) we find a QSH insulator (the non-disordered case corresponds to the point D of (f) in Fig.~2 of the main text) with quantized conductance $2e^2/h$ and vanishing standard deviation in the bulk gap window ($|E_F|<29$ meV). In contrast to the 3QLs case, disorder does not induce any TPTs. This can be understood by the disorder amounting to a negative contribution to the topological mass term within the Born approximation (which is originally negative for 4QLs case). This can be found by using the solution of the closed form of Born approximations after neglecting the self-consistency. The phase boundary between the QSH and the metallic region is well-described by the self-consistent Born approximation while the one between QSH and the Anderson insulator again cannot be obtained by this approximation. 
Even though panels (c) and (d) show a result similar at face value, it demonstrates the existence of a spin-Chern insulator at finite exchange field $gM$. This topological state is not altered by weak disorder and slightly widens in the phase diagram up to around $W \approx 300$ meV as disorder is increased. As we discussed in (a), (b), the disorder does not change the already negative topological mass term to be positive, meaning that once a band inversion occurred, its band does not return to be topologically trivial. The boundary of the spin-Chern insulator and the metal can be explained with the self-consistent Born approximation once more, and when the disorder is increased further, as above, an Anderson insulator is found.
Finally, in the panels (e) and (f) the non-disordered state corresponds to a QAH insulator around $E_F=0$ and we find a similar behavior as in the 3QLs case (compare panels (e) and (f) of Fig.~3 of the main text). The only difference here is the mechanism of how the disorder induces a spin-Chern insulator. Here, at zero disorder the inverted bands from the upper block Hamiltonian was first made trivial by the effects of the exchange field, and upon increasing disorder one recovers its topological nature ($W_{c} \approx 210$ meV). Since there is no time reversal symmetry due to the finite exchange field, the spin-Chern insulator is re-instantiated as is the case for the panels (c), (d) and (e), (f) of Fig.~3 of the main text. Further increasing the disorder drives the system to an Anderson insulator. The QAH to spin-Chern insulator transition is described well by the self-consistent Born approximation.

%\section{Leads for Landauer-B\"uttiker formula}
\section{Influence of the Choice of Leads}

\begin{figure}[t]
{\includegraphics[width=0.95\columnwidth,clip]{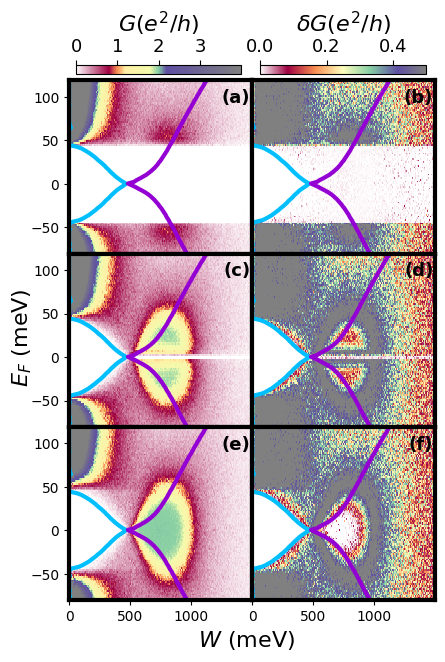}}
\caption{Left (a), (c), (e) and right (b), (d), (f) panels show the average conductance $G$ and the corresponding standard deviation $\delta G$ as a function of disorder strength $W$ and Fermi Energy $E_F$ for 3QLs TI thin film without magnetic doping. Lead types are (a), (b) lead of same type as in the central region, but in the clean limit, (c), (d) metallic lead and (e), (f) QSH lead.The exchange fields $gM$ are taken to be 0 meV. The other parameters and the meaning of the solid and dashed lines are shown in Ref.~\cite{Note_parameter}. %the same as in Fig. 3 of the main text. 
In all calculations, the system sizes are set to $L_x=400a$ and $L_y=100a$ and averages over 50 random configuration are performed. }
\label{3QLsup0}
\end{figure}

\begin{figure}[t]
{\includegraphics[width=0.95\columnwidth,clip]{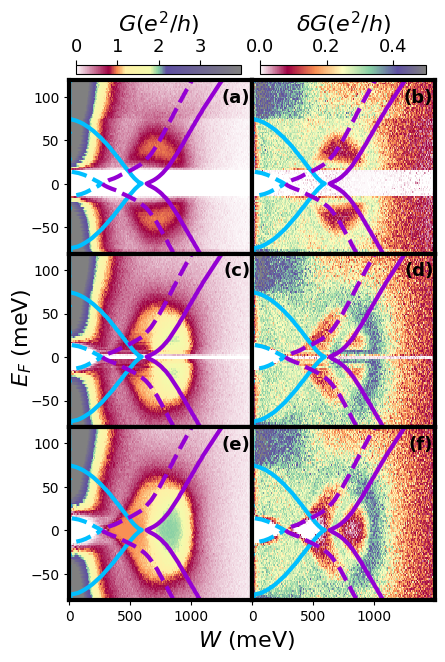}}
\caption{Left (a), (c), (e) and right (b), (d), (f) panels show the average conductance $G$ and the corresponding standard deviation $\delta G$ as a function of disorder strength $W$ and Fermi Energy $E_F$ for 3QLs TI thin films with  doping. Lead types are (a), (b) lead of same type as in the central region, but in the clean limit, (c), (d) metallic lead and (e), (f) QSH lead. The exchange fields $gM$ are taken to be 30 meV. The other parameters and the meaning of the solid and dashed lines are shown in Ref.~\cite{Note_parameter}. %the same as in Fig. 3 of the main text. 
In all calculations, the system sizes are set to $L_x=400a$ and $L_y=100a$ and averages over 50 random configuration are performed.}
\label{3QLsup15}
\end{figure}

\begin{figure}[t]
{\includegraphics[width=0.95\columnwidth,clip]{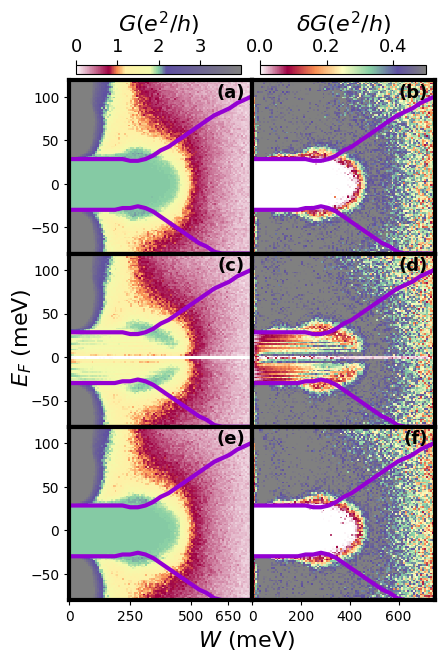}}
\caption{Left (a), (c), (e) and right (b), (d), (f) panels show the average conductance $G$ and the corresponding standard deviation $\delta G$ as a function of disorder strength $W$ and Fermi Energy $E_F$ for 4QL TI thin film without magnetic doping. Lead types are (a), (b) normal sample lead (c), (d) metallic lead (e), (f) QSH lead. The exchange fields $gM$ are taken to be 0 meV. The other parameters and the meaning of the solid and dashed lines are shown in Ref.~\cite{Note_parameter}. %the same as in Fig.~\ref{4QL}. 
In all calculations, the system sizes are set to $L_x=400a$ and $L_y=100a$ and averages over 50 random configuration are performed.}
\label{4QLsup0}
\end{figure}

\begin{figure}[t]
{\includegraphics[width=0.95\columnwidth,clip]{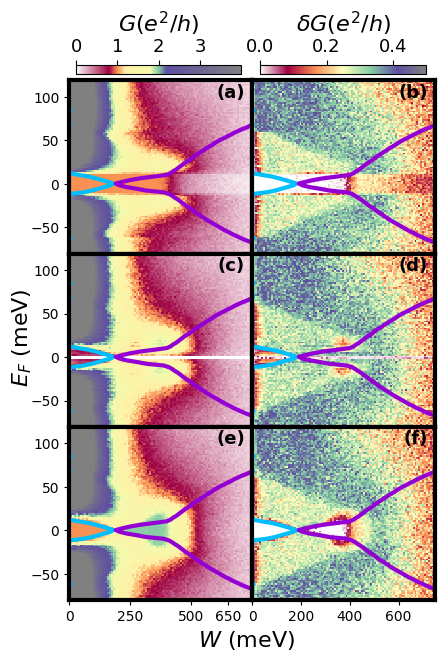}}
\caption{Left (a), (c), (e) and right (b), (d), (f) panels show the average conductance $G$ and the corresponding standard deviation $\delta G$ as a function of disorder strength $W$ and Fermi Energy $E_F$ for 4QLs TI thin film with magnetic doping. Lead types are (a), (b) lead of same type as in the central region, but in the clean limit, (c), (d) metallic lead and (e), (f) QSH lead. The exchange fields $gM$ are taken to be 40 meV. The other parameters and the meaning of the solid and dashed lines are shown in Ref.~\cite{Note_parameter}. %are the same as in Fig.~\ref{4QL}. 
In all calculations, the system sizes are set to $L_x=400a$ and $L_y=100a$ and averages over 50 random configuration are performed.}
\label{4QLsup20}
\end{figure}

Here, we explicitly discuss the influence of choosing different leads on the calculated transport properties shown in the main text, where the lead was chosen in the QSH phase. The main reason why we use the QSH lead is to probe the disorder-induced TPT in the central region most clearly. Using a lead with a trivial bulk gap (such as the clean TI thin film with the thickness of 3QLs), the system is blind to probe the conductance inside the energy window of the band gap, which is the most interesting region to observe the new physics. This is trivial as, since without doping, the Fermi energy $E_F=0$ is firmly inside of the leads bulk gap and therefore we would not be able to detect the new disorder induced topological phases in the central region. Alternatively, we need to dope the lead to be metallic. Therefore, we also %{\color{red}compare to} the transport properties calculated for such a metallic lead. 
compare the transport properties calculated for QSH lead with those for such a metallic lead.  
The latter might be connected closer to the current experimental setup routinely realized. In the metallic case, we can probe the transport signal inside the band gap of the TI thin films and the same physics as shown in the case of the QSH lead can be found. However, due to residual hybridization between the lead and central region, the physics shows up even more clearly with the QSH lead, which we preferred to choose. %our preferred choice of the QSH lead. 

In the following part, we compare the transport properties probed by using these three different kinds of leads: (1) lead of the same nature as the central region but in the clean limit, (2) metallic lead, and (3) QSH lead. %Type (1) is defined as using the identical clean material sample which is intrinsic version of TI thin films in the central region, but without disorder. 
Type (1) is defined as using the disorder-free magnetically doped TI thin films which is the same material used for the central region (without disorder, namely $W=0$). The mass of the left and right lead region $m^{u/l}_{0, L/R}$ is set equal to the the central one $m^{u/l}_{0, C}$. %=m_{0}^{u/l}$., where $m^{u/l}_{0, l/c}$ is the mass term $m^{u/l}_{0}$ in the lead/central ( $m_{0}^{u/l}=m_{0} \pm gM$ as defined in the main text). %{\color{red}{(I am totally confused about the expression. u/l stands for upper/lower block, is that right? $m_0$ is intrinsic mass term, so it should be same for u/l, why do you label it?)}}{\color{blue}{(since there is magnetic doping in the leads as well, we have to distinguish $m^{u/l}_{0}$)}} 
Such setups have been used in  many previous works~\cite{Li2009, Jiang2009}. In the cases of type (2) and (3), we do not consider any disorder and magnetic doping in the leads. %({\color{red}{Takuya: Please double check this expression. In type 2, the mass tern is zero. But in type3, you show $m^{u/l}_{0, l}$=$-|m^{u}_{0, c}|=-|m_{0} + gM|$, you do consider the $gM$. Please double check which part is wrong???, I guess the following part marked by red is wrong}}). 
We define the type (2) lead by setting $m^{u/l}_{0, L/R}=0$, where the lead is a semi-metal. For a type (3) lead, we keep all parameters in the intrinsic TI thin film unchanged except the mass term. In order for the lead to be a QSH insulator with the same bulk gap as the original material used in the central region (without disorder, namely $W= 0$), we always define the sign of the mass term to be negative, namely $m_{0, L/R}$=$-|m^{u}_{0, C}|$ for $|m^{u}_{0, C}|<|m^{l}_{0, C}|$ and $m_{0, L/R}$=$-|m^{l}_{0, C}|$ for $|m^{l}_{0, C}|<|m^{u}_{0, C}|$. The bulk gap value of the QSH leads in the above two cases depends on the magnitudes of $|m^{u}_{0, C}|$ and $|m^{l}_{0, C}|$.
%is simply for the purpose of choosing smaller gap of magnetically doped TI thin film used for the central. 
So in this case within the lead, the helical edge channel always exist ($G=2 e^2/h$), which can connect to edge states and bulk metallic states in the central regions. %{\color{blue}To Peizhe, Ok, the description of Fig. 2 of the main text is unnecessary and makes confusion. So, I remove it. Since we define (2)(3) leads as the one without any disorder and magnetic doping, an additional explanation is unnecessary. Also, I just removed the $u/l$ indices of $m_{0, l}$ since again type (2)(3) leads do not have any disorder and magnetic doping and $u/l$ indices are confusing.}
%In the case of taking the exchange field $gM$ as a variable of the phase diagram, as shown in Fig. 2 of the main text, we use $m^_{0, l}$=$-|m_{0, c}|$ in order to keep leads to be QSH otherwise the leads also would have a phase transition. 
%(I am confused how do you set. The QSH leads are different for Figs.2, 3 and 4 in maintext???)} 
%{\color{red}{(In the lead, do you consider the exchange field or not??? I guess the answer is no. The expression is very misleading, please keep them consistent with those in main text.)}}{\color{blue}{(No, there is no magnetic doping.)}}

%the vicinity of $E_F=0$ which is in the very middle of the gap, the choice of clean normal insulator as a lead does not allow us to detect edge state conductance.
%, which is especially the case for 3QLs, originally normal insulator. 

%the clean Hamiltonian in the 3QLs case

In Figs.~\ref{3QLsup0} and \ref{3QLsup15}, we show the phase diagram of 3QLs with a choice of several different leads. Figures \ref{3QLsup0} (a), (b) and \ref{3QLsup15} (a), (b) clearly show that the conductance vanishes in the energy window of bulk gap and we cannot observe TPTs with the choice of type (1) as a lead. 
%Comparing results shown in Figs.~\ref{3QLsup0} (c-f), we cannot observe gap closing and further band inversion happening inside the bulk gap of central region cannot be seen if normal sample is chosen as leads. 
This is because there is no charge transport from the left lead to the right within the gap energy window even if the central region is metallic. This effect has already been mentioned briefly in Ref.~\cite{Groth2009}, where highly doped leads were used to avoid suffering from this blind spot (compare to Fig. 1 in Ref.~\cite{Groth2009} and Fig. 2(f) in Ref.~\cite{Li2009}).

Although with the choice of metallic lead, we can observe the disorder induced TPTs inside the band gap of TI thin films shown in Figs. \ref{3QLsup0} (c), (d) and \ref{3QLsup15} (c), (d), their edge state conductance obtained with metallic leads is not as clear as QSH leads. In Figs. \ref{3QLsup0} (e), (f) and \ref{3QLsup15} (e), (f), where a QSH lead is employed, clear signatures of edge state transport ($G=2e^2/h$ or $e^2/h$, and vanishing standard deviation $\delta G$ as discussed in the main text) can be observed. %in Figs. \ref{3QLsup0} (e-f) and \ref{3QLsup15} (e-f) while it cannot be seen in Figs. \ref{3QLsup0} (c-d) and \ref{3QLsup15} (c-d).
The reason may lie in interface effect between the semi-metallic lead and the central region. The work about this interface effect and choice of leads will be systematically studied in a future paper. Even though in Fig. \ref{3QLsup15} (f), the standard deviation corresponding to the spin-Chern insulator is not exactly vanishing, this is a finite size effect. This can be clearly seen by comparing Fig. \ref{3QLsup15} (f) with Fig. 3 (d) in the main text, where the former uses $Ny=100$ whereas the latter uses $Ny=200$. %The almost the same discussion can be employed in Figure \ref{3QLsup15}. 
%Because the standard deviation at the Fermi level in this case is almost vanishing , while it will not happens in the case with metallic lead. 
%since the standard deviation is almost vanishing in (e)(f) while it is not the case (c)(d).

In Fig. \ref{4QLsup0}, we show the transport properties of 4QLs thin films comparing the influence of choosing different leads. Because, by definition, case (1) is also a QSH lead, cases (1) and (3) as choices of leads are equivalent and consequently the same results are observed in Figs. \ref{4QLsup0} (a), (b) and (e), (f). %{\color{red}In the case of a metallic lead, shown in Figs. \ref{4QLsup0} (c), (d), disorder stabilized the QSH phase, which is clearly detected in the TI thin film bulk gap region for QSH leads (see Figs. \ref{4QLsup0} (a) ,(b) and (e), (f)), cannot be obtained due to the interface effect clearly using metallic leads.}
In the case of a metallic lead, shown in Figs. \ref{4QLsup0} (c), (d), the QSH phase clearly detected inside the bulk gap %in the TI thin film bulk gap region 
for QSH leads (see Figs. \ref{4QLsup0} (a) ,(b) and (e), (f)), cannot be clearly obtained due to the interface effect using metallic leads.

In the case of finite exchange field as shown in Fig.~\ref{4QLsup20}, corresponding to QAH phase without disorder, we can see the same phenomena discussed in Figs.~\ref{3QLsup0} and \ref{3QLsup15}. With the choice (1) for a lead, we cannot clearly detect the TPTs from the QAH to the spin-Chern insulator because the largest current of the lead ($G=e^2/h$) is smaller than the edge conductance of spin-Chern insulator ($G=2e^2/h$). %(Are these expressions right?? In such case, the normal lead is a QAH???)} %{\color{blue}{(yes. I define normal material including magnetic doping)}} %Thus, inside the gap the conductance cannot be more than $G=e^2/h$. 
However, since the QSH lead hosts the edge current of $G=2e^2/h$, %to the central in the bulk gap energy window in (e)(f), 
we can observe the TPT induced by disorder from a QAH to spin-Chern insulator in the central region when using a QSH lead. Although, for metallic leads similar trait can be seen in Figs.~\ref{4QLsup20} (c), (d) as in (e), (f), the signature of edge state conductance is not as clear as in the case of QSH leads again due to interface effect.

%(B/max(EF^2-m0^2))(pi/a)^2~3.0...>1
%
%Fig 2 (a) and (b)
%(a) and (b) of Fig 3
%At large exchange fields a transition to a QAH phase is found for both 3QLs and 4QLs, either by inverting
%in contrast
%We start the discussion with
%Further enhancing the disorder potential
%shown in panels (c) and (d)
%to drive the thin film to be a Cher
%be well understood by the self-consistent Born approximation.
% disorder tunes
%e as disorder is turned o
%are captured well within the self-consistent Born approximation at small disorder
%The boundary of the spin-Chern insulator and the metal can be explained with the self-consistent Born approximation once more, and when the disorder is increased further, as above, an Anderson insulator is found.

%\bibliography{apssamp}
\bibliography{Manuscript}% 